\documentclass[12pt]{article}
\usepackage{epsfig, amssymb}
\usepackage{graphicx,epsfig}
\begin{document}
\begin{titlepage}
\thispagestyle{empty}
\begin{flushright}
\end{flushright}

\bigskip

\begin{center}
\noindent{\Large \textbf
{Boundary conditions for $SU(2)$ Yang-Mills on $AdS_4$}}\\

\vspace{2cm} \noindent{Dileep P. Jatkar${}^{a}$\footnote{e-mail:dileep@hri.res.in} and
Jae-Hyuk Oh${}^{a}$\footnote{e-mail:jack.jaehyuk.oh@gmail.com}}

\vspace{1cm}
{\it
Harish-Chandra Research Institute, \\
  Chhatnag Road, Jhunsi, Allahabad-211019, India${}^{a}$\\
}
\end{center}

\vspace{0.3cm}
\begin{abstract}
We consider $SU(2)$ Yang-Mills theory on $AdS_4$ 
  %% to construct its boundary $CFT$(also its dual $CFT$ i.e. its
  %% effective action)
  by imposing various boundary conditions, which correspond to
  non-trivial deformations of its boundary $CFT$.  We obtain 
  classical solutions of Yang-Mills fields up to the first subleading
  order correction by using small amplitude expansion of the gauge
  field 
%%   power expansion order by order in small
%%   amplitude to retain the leading gauge interaction 
  without
  considering gravitational back reaction. We also consider $SU(2)$
  Yang-Mills instanton solution in $AdS_4$ bulk, and propose
  a boundary action.  It turns out that the boundary theory
  is the Chern Simons theory with a non-local deformation which has the
  form similar to the Wilson line.
%%   , (Wilson loops
%%   coupling to external sources). 
  In the limit of the deformation parameter $\rho \to \infty$, this non-local
  deformation is suppressed and the boundary theory becomes pure
  Chern Simons.  For large but finite values of $\rho$, this non-local
  deformation can be treated perturbatively within the Chern-Simon
  theory.
%%   We have considered $SU(2)$ Yang-Mills theory on $AdS_4$ to construct
%%   its boundary $CFT$(also its dual $CFT$ i.e. its effective action) by
%%   imposing various boundary conditions, which correspond to non-trivial
%%   deformations to its boundary $CFT$. We have obtained the classical
%%   solutions of Yang-Mills fields up to the first subleading order
%%   correction by power expansion order by order in small amplitude to
%%   retain the leading gauge interaction without considering
%%   gravitational back reaction. We also consider $SU(2)$ Yang-Mills
%%   instanton solution in $AdS_4$ bulk, and construct its reasonable
%%   boundary action.  It turns out that the boundary theory is Chern
%%   Simons theory with a non-local deformation, (Wilson loops coupling
%%   to external sources). In some limit, this non-local deformation can
%%   be suppressed and the boundary theory becomes pure Chern Simons.
  % It is discussed that under suitable boundary condition, bulk dual
  % theory can be reconstructed by boundary theory via Stochastic
  % Quantization. We have conjectured that cut-off independent counter
  % term action would be a natural candidate to construct
  % Fokker-Planck action for the right bulk theory reconstruction and
  % we have demonstrated some examples.
\end{abstract}
\end{titlepage}

\newpage

\tableofcontents
%\newpage

\section{Introduction}
``Alternate quantization'' was first studied by Breitenlohner and
Freedman\cite{Freedman1} in the context of compactifications of
supergravity theories to anti-de Sitter space. In the wake of
developments in AdS/CFT correspondence, there has been a renewed
interest in it.  Klebanov and Witten\cite{Witten1} first discussed it
in AdS/CFT, providing interesting boundary conformal field theories
generated by the alternate scheme\cite{Witten3,Amit1}.  In a certain
window of the conformal dimensions of the boundary composite operators
corresponding the bulk excitations in $AdS_{d+1}$, there are two
possible quantization schemes for their boundary $CFT$s.

For massive scalar fields in $AdS_{d+1}$, in the window of their
masses as $ -\frac{d^2}{4}\leq m^2 \leq -\frac{d^2}{4}+1$, there are
two possible quantization schemes, so two possible boundary conformal
field theories, which are called $\Delta_+$-theory and
$\Delta_-$-theory, where
$\Delta_{\pm}=\frac{d}{2}\pm\sqrt{\frac{d^2}{4}+m^2}$, which is
conformal dimension of the boundary operator in each theory
\cite{Witten1}.  In this window, both $CFT$s are above the
unitarity bound, $\Delta_\pm \geq \frac{d}{2}-1$, so their two point
correlators are positive definite in the position space.  In
$\Delta_+$-theory, near $AdS$ boundary expansion of the bulk scalar
fields is $\phi(r,x^\mu)|_{r \rightarrow
  0}=\phi_{0}(x^\mu)r^{d-\Delta_{+}}+A(x^\mu)r^{\Delta_+}$.  It is
well known that $A(x^\mu)$ corresponds to certain composite operator
in the $CFT$ and $\phi_{0}(x^\mu)$ corresponds to the external source
coupled to it. In fact, $\Delta_+$-theory is not independent from
$\Delta_-$-theory.  They are related to each other by Legendre
transform.  This Legendre transform switches the role of $A(x^\mu)$
and $\phi_{0}(x^\mu)$ in $\Delta_-$-theory, because they are
canonical conjugates of each other.

In the dual gravity theory, two possible boundary $CFT$s can be
obtained by imposing different boundary conditions.  Boundary
conditions for bulk fields and corresponding boundary terms in the
AdS/CFT context has been studied by various authors in the past.  For
Dirac fields it was studied by Henneaux \cite{Henneaux:1998ch},
whereas for Rarita-Schwinger fields it was analysed in
\cite{Rashkov:1999ji}.  This issue in case of inequivalent
quantization was addressed by \cite{Witten3,Mueck:1999kk} and in
Lorentzian AdS/CFT case was dealt with in \cite{Marolf:2004fy}.  For
$\Delta_+$-theory, the corresponding boundary condition is the
Dirichlet boundary condition as $\delta \phi_0(x^\mu)=0$. Dirichlet
boundary condition is the usual boundary condition in AdS/CFT context.
Since the $\Delta_+$-theory is unitary even when $m^2 \geq
-\frac{d^2}{4}+1$, Dirichlet boundary condition is always a possible
boundary condition.  The $\Delta_-$-theory can be obtained by imposing
Neumann boundary condition, $\delta A(x^\mu)=0$. This Neumann boundary
condition is obtained by adding boundary term $S_{bdy}\sim\int
\phi(x^\mu)A(x^\mu) d^dx$ at $r=0$.  Adding such a term in turn
generates the same effect as performing the Legendre transform of the
$\Delta_+$-theory, therefore such a boundary term takes the
$\Delta_+$-theory to the $\Delta_-$-theory.

The Neumann boundary condition can be generalized by deforming the
boundary $CFT$ by adding a general form of $S_{bdy}$.  Such a boundary
action can be an arbitrary function of $\phi_0(x^\mu)$ and $A(x^\mu)$.
By adding the boundary action, one can obtain the ``on-shell action''
$I_{os}=S_{bulk}+S_{bdy}$, where $S_{bulk}$ is the boundary
contributions from the bulk action.  The boundary condition is
obtained by performing functional variation of the on-shell action and
setting it to zero, $\delta I_{os}=0$.  This corresponds to saddle
point of the on-shell action, which is the classical vacuum of the
boundary theory.

An interesting example with general deformations is the conformally
coupled scalar field theory in $AdS_4$\cite{Sebastian2}.  In
\cite{Sebastian2} authors consider a massless scalar field theory with
$\lambda\phi^4$ and $\frac{1}{6}R\phi^2$ interactions, where $\phi$ is
the scalar field, $\lambda$ is the quartic self-coupling of the scalar
field and $R$ is curvature scalar of $AdS_4$.  The boundary theory
corresponding to the conformally coupled bulk scalar field contains a
triple trace deformation term $S_{bdy} \sim \alpha \int d^3x
\phi^3_{0}(x^\mu)$, where $\alpha$ is a numerical parameter and
$\phi_{0}$ is the boundary value of the scalar field $\phi$.  Under
the field redefinition $\phi_0=\varphi^2$ and truncation up to the
second order in small derivative expansion, the boundary on-shell
action takes the canonical form with $\varphi^6$ coupling\cite{Sebastian3}.
%% This boundary expanding in small derivatives up to two-derivatives.

$\Delta_\pm$-theories for $U(1)$ vector fields in $AdS_4$ are
well-defined \cite{Witten2,Donald1,Sebastian1} in which $\Delta_+=2$
and $\Delta_-=1$. The unitarity bound for the vector like local
observables in $d$-dimensional $CFT$ is $\Delta \geq
d-1$\cite{Minwalla1}, so $\Delta_+=2$ theory is unitary when $d=3$.
It also would imply that the $\Delta_-=1$ theory does not satisfy the
unitarity bound.  One way to interpret the dual operator with
conformal dimension `1' (the conjugate of source term from the bulk
action) is as the $U(1)$ vector gauge field in the boundary theory.
Clearly there is an ambiguity in defining this operator due to
non-invariance of it under gauge transformations but instead one
interprets it as an observable which is not local\cite{Leigh1}.  The
field strength constructed out of this gauge fields is in fact a local
observable.  It also resolves the apparent contradiction with the
unitarity bound because the field strength has conformal dimension 2,
which satisfies the unitarity bound.  There are many possible boundary
deformations which provide interesting on-shell actions. In
\cite{Sebastian1}, the authors consider ``massive deformation'' from
which they derive the on-shell action to be the massive gauge field
action. In the case that self-duality condition together with massive
deformation, one obtain massive Chern Simons boundary on-shell
action\cite{Townsend1}.

In this paper, we extend the discussion to non-abelian gauge theory in
$AdS_4$. Before summarizing our main result, let us briefly explain
our motivations.  The motivations are three folds. First, Dirichlet
and Neumann boundary conditions in abelian gauge field theory in the
bulk correspond to free $CFT$. The most natural way to introduce
interactions is to consider Yang-Mills theory.  As will be shown, this
will give non-trivial momentum dependent interactions in the boundary
action for Dirichlet and Neumann boundary conditions.  We will discuss
generalizations of other deformations which in the abelian context
were considered in \cite{Sebastian1}.  
Second, it is well-known that abelian gauge field theory action in 4-dimension is manifestly invariant under electric-magnetic duality, 
and it is also successfully embodied in $AdS$ space\cite{Sebastian1}.
We want to extend this duality to Yang-Mills. In \cite{Deser2}, it is reported that when one retains cubic order interactions only, one can implement
electric-magnetic duality in Yang-Mills system upto that order, even if that is not possible to construct with quartic interaction terms. 
In fact, this symmetry is not manifest 
electric-magnetic duality since it turns out that the variation of electric field is not proportional to magnetic field, but it is the most natural extension of 
the abelian duality. We will discuss how this symmetry is embodied in $AdS$ space.
Finally, Yang-Mills theory on
$AdS_4$($U(1)$ gauge theory too) should be the same with that in the
half of 4-dimension flat space through Weyl scaling of the $AdS_4$
metric.  Therefore, $SU(2)$ Yang-Mills instanton solution in $\mathbb R^4$ can
easily be adapted for $AdS_4$ as well. As $r \rightarrow 0$,
Yang-Mills instanton has non-trivial boundary value whereas near
Poincare horizon $r\to\infty$ it becomes a pure gauge
solution. Therefore, Yang-Mills instanton definitely changes the
boundary condition on the $AdS_4$ boundary and exploring implications
of these boundary condition is interesting.

In our study, we develop boundary deformations with certain reasonable
boundary conditions derived from perturbative and non-perturbative
bulk solutions. We briefly list the results here. For perturbative
approach, we solve bulk Yang-Mills equations of motion in power
expansion order by order in small amplitudes of Yang-Mills fields. To
retain the leading interactions in Yang-Mills coupling $g$, we obtain
the bulk solutions up to first subleading order corrections in the
small amplitudes. Up to this order, we can account for cubic
interactions in the boundary on-shell actions(also its dual $CFT$
actions). 
%To get boundary on-shell actions, $I_{os}$ with various
%boundary conditions, we deform the on-shell action with various
%boundary actions $S_{bdy}$, where $I_{os}=S_{bulk}+S_{bdy}$ and
%$S_{bulk}$ is the boundary contributions from the bulk action. 
In the Dirichlet boundary condition case(in which case, the boundary source
becomes $A^{a(0)}_i$, the boundary value of Yang-Mills fields), the
boundary on-shell action $I^{D}_{os}$ gives rise to the 
propagator which is proportional to absolute value of the 3-momenta,
$|q|$, and exotic 3-momenta dependent cubic interactions as
\begin{equation}
  \Delta_{ijk}^{D,abc}(q,l,p) \sim ig\epsilon^{abc}\delta^{3}(q+p+l)
  \frac{(l-q)_k\delta_{ij}+(p-l)_i\delta_{jk}+(q-p)_j\delta_{ik}}{2(|q|+|p|+|l|)},
\end{equation}
where $\Delta^{D,abc}_{ijk}(q,l,p)$ are 3-point function on the
boundary $CFT$ $a$,$b$,$c$ are $SU(2)$ gauge indices and $q_i$, $l_i$
and $p_i$ are 3-momenta along the boundary directions with the
boundary spacetime indices $i,j,k=1,2,3$. For the Neumann boundary
condition(the source becomes $-A^{a(1)}_i$, which is the boundary
value of the canonical momentum of Yang-Mills fields), the 
propagator is proportional to $\frac{1}{|q|}$ and the 3-point function
is given by
\begin{equation}
\Delta_{ijk}^{N,abc}(q,l,p) \sim \frac{ \Delta_{ijk}^{D,abc}(q,l,p)}{|q||p||l|} .
\end{equation}
The most interesting cases are massive and self-dual boundary
conditions. The massive boundary condition is written as
\begin{equation}
(\sqrt{-\nabla^2}-m)A^{a(0)}_i(x^i)=0,
\end{equation}
where $m$ is a mass dimension 1 constant. Once one applies this
boundary condition to the bulk fields, then the boundary theory
becomes massive gauge field theory.  In case of the self-dual boundary
condition we do small $r$ expansion of the  self-duality condition,
\begin{equation}
F^{a}_{MN}=\frac{1}{2}\epsilon_{MNPQ}F^a_{PQ},
\end{equation}
which is given by
\begin{equation}
A^{a(1)}_{i}=\frac{1}{2}\epsilon_{ijk}F^{a}_{jk}.
\end{equation}
Imposing these two different boundary conditions on the bulk fields,
we get non-abelian massive Chern Simons gauge theory on the
3-dimensional boundary.

We also explore the non-perturbative solutions from the bulk
theory. We consider Yang-Mills instanton with its winding number 1. It
turns out that one of the possible boundary conditions that we can
impose results in the boundary term consisting of the Chern Simon
action with some non-local deformations.  In this non-local
deformation contains a line integration of the form
\begin{equation}
 \sim e^{\int^z_0\epsilon^{ai}_{\ \ j} A^{a(0)}_i(\tilde z) d\tilde z^j}.
\end{equation}
This kind of non-local interactions never comes out from any
perturbative deformations, which can be the genuine properties from
the bulk instanton backgrounds.

This note is organized as follows. In Sec.\ref{Bulk Solutions}, the
bulk equations of motion of Yang-Mills fields are solved and its
perturbative solutions are obtained up to the first subleading order
corrections in the small amplitudes of Yang-Mills fields.  In
Sec.\ref{Boundary Conditions and the Effective Action}, we perform 
various boundary deformations of the on-shell actions and obtain some
interesting boundary actions.  In Sec.\ref{em-duality}, we explore
implications of approximate electric-magnetic duality in the context
of $SU(2)$ Yang-Mills theory in $AdS_4$.  In Sec.\ref{Yang-Mills
  Instanton}, we explore Yang-Mills instanton solutions and their
boundary conditions.  In the conclusion section we summarise our
results and discuss their implications.  Some technical details
are presented in appendices.

\section{$SU(2)$ Yang-Mills on $AdS_4$ and its Solution}
\label{Bulk Solutions}\setcounter{equation}{0}
We start with the $SU(2)$ Yang-Mills(Euclidean) action in $AdS_4$
space-time background as
\begin{equation}
S[A]=\frac{1}{4}\int d^4x \sqrt{G} F^a_{MN}F^{aMN},
\end{equation}
where the space-time indices $M,N$ run from 1 to 4 and the gauge
indices $a$ does from 1 to 3.  The background metric, $G$ is
\begin{equation}
ds^2=G_{MN}dx^Mdx^N=\frac{dr^2+\sum_{i=1}^{3}dx^i dx^i}{r^2},
\end{equation}
where we define the radial coordinate $r$ as $r \equiv x^4$, the
indices $i,j..$ are defined boundary space-time coordinate, which run
from 1 to 3.  Yang-Mills field strength is given by
\begin{equation}
F^{a}_{MN}=\partial_{M} A^a_{N}-\partial_{N} A^a_{M}-g\epsilon^{abc}A^b_{M} A^c_N.
\end{equation}

One interesting feature of this system is that the Weyl rescaling of
background metric, $ds^2 \rightarrow r^2 ds^2$, maps the Yang-Mills
theory in $AdS_4$ to that defined in 4-dimensional flat space-time.
The space-time is a half of $\mathbb R^4$, because the radial
coordinate in $AdS$ space runs from $0$ to $\infty$. Therefore, the
Yang-Mills action becomes
\begin{equation}
\label{flat-space-action}
S[A]=\frac{1}{4}\int_{\mathbb R^4_{+}} d^4x  F^a_{MN}F^{aMN},
\end{equation}
where the space-time indices are contracted with $\delta_{MN}$ and
$\mathbb R^4_{+}$ denotes a half of the 4-dimensional flat space.
% \section{Bulk Solutions}
% \label{Bulk Solutions}

In this section, we evaluate bulk equations of motion and obtain their
solutions with a power series expansion in small amplitude of
Yang-Mills fields(This expansion is effectively the same as the
Yang-Mills coupling $g$ expansion).  We would solve Yang-Mills
equations up to the first sub-leading order corrections to take into
account effects of interaction terms in Yang-Mills action.  Up to this
order, only cubic interactions terms participate.  The equations of
motion are given by
\begin{equation}
\label{equation-of-motion}
0=\mathcal D_M F^{a}_{MN}=\partial_M  F^a_{MN}+g\epsilon^{abc} F^{b}_{MN}A^{c}_{M},
\end{equation}
where $\mathcal D_{M}$ is the gauge covariant derivative.  To evaluate
the perturbative equations of motion, we expand Yang-Mills field as
\begin{equation}
A^{a}_M=\varepsilon \bar A^a_M+\varepsilon^2 \tilde A^a_M+O(\varepsilon^3).
\end{equation}
where $\varepsilon$ is a book keeping parameter for the expansion,
which is a dimensionless small number.  The equations of motion are
evaluated for each order in $\varepsilon$ as
\begin{eqnarray}
\label{first-order-EOM}
{\rm First\ Order\ :\ }0&=&\partial_{M}(\partial_{M}\bar{A}^a_N-
\partial_{N}\bar{A}^a_M), \\
\label{scond-order-EOM}
{\rm Second\ Order\ :\ }0&=&\partial_{M}(\partial_{M}\tilde{A}^a_N-
\partial_{N}\tilde{A}^a_M)-g\epsilon^{abc}\left( \partial_M(\bar A^b_M 
\bar A^c_N) \right. \\ \nonumber
&&-\left. (\partial_M\bar{A}^b_N-\partial_N\bar A^b_M)\bar A^c_M \right),
\end{eqnarray}
and so on.
% If one keeps $\varepsilon \bar A^a_M$ only, this becomes a theory
% with 3-copies of abelian gauge fields.
We start with the first order equations in $O(\varepsilon)$ are given by
\begin{eqnarray}
0&=&\nabla^2 \bar A^a_{r} -\partial_r \partial_i \bar A^a_i, \\
0&=&(\partial^2_r +\nabla^2)\bar A^a_i-\partial_{i}(\partial_r 
\bar A^a_{r}+\partial_j \bar A^a_j),
\end{eqnarray}
where we split the indices $M,N..$ into $r$ and $i,j..$ and $\nabla^2
\equiv \sum_{j=1}^3 \partial_j \partial_j$.  At this order, equations
of motion are identical to 3 copies of $U(1)$ gauge theory equations.
Solutions to these equations has already been obtained in
\cite{Sebastian1} (See Sec.2 and Appendix.B in it). We briefly list
the leading order solution $\bar A^a_M$, in momentum space as
\begin{eqnarray}
\label{first order solution}
\bar A^a_{i,q}(r)&=&\bar A^{aT}_{i,q}(r) -iq_i \bar\phi^a_q(r) {, \  \
} \bar A^{a}_{r,q}(r)=\partial_r \bar\phi^a_q(r){, \  \  } q_i 
\bar A^{aT}_{i,q}(r)=0, \\ \nonumber
{\rm \ and \ } \bar A^{aT}_{i,q}(r)&=&\bar A^{aT(0)}_{i,q}\cosh(|q|r)
+\frac{1}{|q|}\bar A^{aT(1)}_{i,q}\sinh(|q|r),
\end{eqnarray}
where $q_i$ are three momenta along the boundary direction and 
the solution is obtained by using Fourier transform of the
position space representation defined as
\begin{eqnarray}
\label{Fourier Transform}
\Phi^a_M (x,r)&=&\int^{\infty}_{-\infty} d^3 p
e^{-ip_{i}x_{i}}\Phi^a_{Mp}(r), \\ \nonumber
\Phi^a_{Mp}(r)&=&\frac{1}{(2\pi)^3}\int^{\infty}_{-\infty} d^3 x 
e^{ip_{i}x_{i}}\Phi^a_M (x,r),
\end{eqnarray}
where $\Phi$ denotes any fields appearing in the bulk theory.  $\bar
A^{aT}_{i,q}$ is the transverse part of the gauge field, which is given by
\begin{equation}
\bar A^{aT}_{i,q} = P_{ij}(q) \bar A^{a}_{j,q},
\end{equation}
where we define a projection operator,
\begin{equation}
P_{ij}(q)=\delta_{ij}-\frac{q_i q_j}{q^2}.
\end{equation}
%and $\bar A^{a(0)}_{i,q}$ and $\bar A^{a(1)}_{i,q}$ are arbitrary vector coefficients
If $\bar A^a_{i,q}(r)$ and $\bar A^a_{r,q}(r)$ are solutions then
\begin{equation}
\bar A^{a\prime}_{i,q}(r)=\bar A^a_{i,q}(r)-iq_i \bar\phi^a_q(r) {\rm
  \ and \ } \bar A^{a\prime}_{r,q}(r)=\partial_r \bar\phi^a_q(r)
\end{equation}
also solve the equations of motion.  $\bar\phi^a$ is a gauge freedom
which is not completely determined by equations of motion.

To proceed further we will use the radial gauge, namely $\bar
A^{a}_{r,q}(r)=0$. In the radial gauge, the residual gauge freedom is
obtained by restricting the gauge parameter $\bar\phi^a_{r,q}(r)$ to be
independent of $r$,
\begin{equation}
\label{the radial gauge}
\bar \phi^a_q(r) \rightarrow \bar\phi^a_q.
\end{equation}
Then by definition, $\bar A^{aT}_{i,q}(r)$ is gauge invariant under this
residual gauge transformation.
%so without loss of any generality one can set $\bar\phi^a=0$.
% The solution is obtained in momentum space using Fourier transform as
% \begin{eqnarray}
% \label{Fourier Transform}
% \Phi^a_M (x,r)&=&\int^{\infty}_{-\infty} d^3 p e^{-ip_{i}x_{i}}\Phi^a_{Mp}(r), \\ \nonumber
% \Phi^a_{Mp}(r)&=&\frac{1}{(2\pi)^3}\int^{\infty}_{-\infty} d^3 x e^{ip_{i}x_{i}}\Phi^a_M (x,r).
% \end{eqnarray}
For the regularity of the solutions at the Poincare horizon, at
$r=\infty$, we require that
\begin{equation}
\label{regularity-condition}
\bar A^{aT(0)}_{ip}+\frac{1}{|p|}\bar A^{aT(1)}_{ip}=0.
\end{equation}
This removes the term proportional to $e^{|p|r}$ near the Poincare horizon.
Using this regularity condition we can write the solution in the following form 
\begin{equation}
\label{first order bulk solution}
\bar A^{aT}_{i,p}(r)
%=\frac{1}{2}\left(A^{aT(0)}_{ip}-\frac{1}{|p|}A^{aT(1)}_{ip}\right) e^{-|p|r}
=\bar A^{aT(0)}_{ip}e^{-|p|r}.
\end{equation}

After having obtained leading order solution, we will now solve the
equation second order in $\varepsilon$. The
precise procedure is given in Appendix.\ref{Evaluation of the Second
  order Bulk Solution}, here we briefly discuss the equations and
their solutions.  The equation for $N=r$ and $N=i$ from
eq.(\ref{scond-order-EOM}) become
\begin{eqnarray}
\label{N=r the second order Bulk equation}
0&=& \left( \nabla^2\tilde A^a_{r}-\partial_r \partial_j 
\tilde A^a_j \right) -g\epsilon^{abc}\left( \partial_i(\bar A^b_i 
\bar A^c_r)-\bar A^c_i
(\partial_i \bar A^b_r-\partial_r \bar A^b_i) \right), \\
\label{N=i the second order Bulk equation}
0&=&(\partial_r^2+\nabla^2)\tilde A^a_i -\partial_i 
(\partial_r \tilde A^a_r+\partial_j \tilde A^a_j)-g\epsilon^{abc}
\left( \partial_r (\bar A^b_r \bar A^c_i)
+\partial_j (\bar A^b_j \bar A^c_i)\right. \\ \nonumber
&-&\left.(\partial_j \bar A^b_i - \partial_i \bar A^b_j)\bar A^c_j 
-(\partial_r \bar A^b_i-\partial_i \bar A^b_r) \bar A^c_r \right),
\end{eqnarray}
respectively. 
% These equations are manifestly gauge covariant
% (The detailed proof is given in Appendix.\ref{Gauge Transformation}).
%To prove it, one can use eq.(\ref{first-order-gauge-transform}), 
% eq.(\ref{second-order-gauge-transform}) and the first order
% EOM in $\varepsilon$, eq.(\ref{first-order-EOM}). 
% With this gauge covariance, one can set the gauge freedom 
% $\tilde \phi^{a}=0$ as the leading order solution
% case($\tilde \phi^{a}$ is defined in Appendix.\ref{Gauge Transformation}).

The first order solution in $\varepsilon$, $\bar A^a_{i,q}$ appear in
the form of source terms in the second order equations in
$\varepsilon$, (\ref{N=r the second order Bulk equation}) and
(\ref{N=i the second order Bulk equation}).  Using this we divide up
the second order solutions into the homogeneous part $\tilde
A^{a}_{i,q(H)}$ and inhomogeneous part $\tilde A^{a}_{i,q(I)}$.
%There are also homogeneous solutions in the second order in $\varepsilon$.
The homogeneous solution has the same form with $\bar A^a_M$ as in
eq.(\ref{first order solution}).  This is because the homogeneous
equations are again linear and are identical to
eq.(\ref{first-order-EOM}).  With the regularity condition in the
interior, we get $\tilde A^{a}_{i,p(H)}(r)=\tilde A^{a(0)}_{i,p(H) }e^{-|p|r}$ and by
imposing the radial gauge, the solution takes the form
\begin{equation}
\label{the second order homogeneous solution}
\tilde A^{a}_{i,p(H)}(r)=\tilde A^{aT(0)}_{ip(H)}e^{-|p|r}-ip_i \tilde \phi^a_{p(H)},
\end{equation}
where as argued previously, $\tilde A^{aT(0)}_{ip(H)}e^{-|p|r}$ is
gauge invariant part of the solutions and $\tilde \phi^a_{p(H)}$ is
gauge parameter, which also does not depend on $r$ like $\bar
\phi^a_{p}$ in the radial gauge.
%where the subscript, $H$, indicates that this solution is a
%homogeneous part of the EOM in $O(\varepsilon^2)$.

With the projection operator, one can split the inhomogeneous part
$\tilde A^a_{i,q(I)}$ into two pieces as
\begin{equation}
\tilde A^{a}_{i,q(I)}=\tilde A^{aT}_{i,q(I)}+\tilde A^{aL}_{i,q(I)},
\end{equation}
where $\tilde A^{aT}_{i,q(I)}=P_{ij}(q)\tilde A^{a}_{j,q(I)}$ is the
transverse part of the gauge field and $\tilde
A^{aL}_{i,q(I)}=\frac{q_iq_j}{q^2}\tilde A^{a}_{j,q(I)}$ is the
longitudinal part of the inhomogeneous
solutions. %(For convenience, we have omitted the subscripts $(I)$ in them).
The equations are also separated into the longitudinal and the transverse part,
which are given by
% eq.(\ref{r-equation}) and eq.(\ref{2nd order transverse solution})
% with that the gauge parameter sets to be zero are given by
\begin{equation}
\label{the final form of longitudinal}
-iq_j \partial_r \tilde A^{aL} _{j,q(I)}(r)
=g\epsilon^{abc}\int^{\infty}_{-\infty} d^3 p |q-p| \left( 
 \bar A^{cT}_{j,p}(r)\bar A^{bT}_{j,{q-p}}(r) 
-iq_j \bar \phi^c_p \partial_r  A^{bT}_{j,q-p}(r)\right),
\end{equation}
\begin{eqnarray}
\label{the final form of TRAnsverse}
(\partial^2_r-q^2)\tilde A^{aT} _{i,q(I)}&=& 
g \epsilon^{abc}\int^{\infty}_{-\infty}d^3p
 \bar A^{cT}_{k,p}(r) \alpha_{ijk}(p,q) \bar A^{bT}_{j,q-p}(r) 
\\ \nonumber
&+&g\epsilon^{abc}P_{ij}(q)\int^\infty_{-\infty}d^3 p ((q-p)^2-q^2)
\bar \phi^b_p \bar A^{cT}_{j,q-p}(r)\\ \nonumber
&-&\frac{i}{2}g\epsilon^{abc}q^2P_{ij}(q)\int^\infty_{-\infty}d^3 p 
p_j\bar \phi^b_p \bar \phi^c_{q-p} 
\end{eqnarray}
in the radial gauge,
where
\begin{equation}
\label{definition of alpha}
\alpha_{ijk}(p,q) \equiv \left( \frac{iq_i}{q^2}(q-p)^2-i(q-p)_i
\right)\delta_{jk}+iq_k\delta_{ij}-iq_j \delta_{ik}.
\end{equation}
% In eq.(\ref{the final form of longitudinal}), one of $\tilde
% A^{a}_{rq}$ or $\tilde A^{aL}_{iq}$ is able to be zero, but not both
% of them.  Thus, one of which is the real degrees of freedom.
% % If one chooses the radial gauge for the second order solutions
% % too, then we must have longitudinal mode and vice versa. (Dileep,
% % do you think it is correct?)
% % \subsection{Boundary conditions and Constraints}
% % We also discuss about gauge choice.
% There are several ways of choosing gauge. One can choose radial
% gauge, $ A^a_r=0$, and another choice is transverse gauge,
% $A^a_i=P_{ij}A^{a}_{j}$(Or $A^{aL}_{i,q}=0$). In our context, we
% choose the radial gauge.  Under such gauge condition,
Solutions to these equations in the momentum space are given by
\begin{eqnarray}
\label{the longitudinal the secon order}
\tilde
A^{aL}_{i,q(I)}(r)&=&-ig\epsilon^{abc}\frac{q_i}{q^2}\int^\infty_{-\infty}
d^3p \bar A^{cT(0)}_{j,p} \bar
A^{bT(0)}_{j,q-p}\frac{|q-p|}{|p|+|q-p|}e^{-(|p|+|q-p|)r} \\ \nonumber
&-&g\epsilon^{abc}\frac{q_i q_j}{q^2}\int^\infty_{-\infty}d^3p 
\bar \phi^c_p \bar A^{bT(0)}_{j,q-p}e^{-|q-p|r} + q_i f^a_q, \\ 
\label{the transverse the secon order}
\tilde A^{aT}_{i,q(I)}&=&g\epsilon^{abc}\int^\infty_{-\infty}d^3p 
\bar A^{cT(0)}_{k,p} \bar
A^{bT(0)}_{j,q-p}\frac{\alpha_{ijk}(p,q)}{(|p|+|q-p|)^2
-q^2}e^{-(|p|+|q-p|)r} \\ \nonumber
&+&\frac{i}{2}g\epsilon^{abc}P_{ij}(q)\int^\infty_{-\infty} d^3 p p_j
\bar \phi^b_p \bar \phi^c_{q-p} + g\epsilon^{abc}
P_{ij}(q)\int^\infty_{-\infty}
d^3 p \bar \phi^b_p \bar A^{cT(0)}_{j,q-p}e^{-|q-p|r},
\end{eqnarray}
where $f^a_q$ is an integration constant.  Therefore, the total
solution up to $O(\varepsilon^2)$ is given by
\begin{equation}
\label{total solutions}
A^{a}_{i,p}(r)=\varepsilon \bar A^{a}_{i,p}(r) + \varepsilon^2 
(\tilde A^{a}_{i,p(H)}(r) + \tilde A^{aT}_{i,p(I)}(r)+ \tilde
A^{aL}_{i,p(I)}(r) )+O(\varepsilon^3),
\end{equation}
under radial gauge. This total solution can be sorted out into gauge
invariant parts and gauge parameter dependent parts. The transverse
parts in $\bar A^a_{i,q}(r)$ and $\tilde A^a_{i,q(H)}(r)$ are gauge
invariant.  In eq.(\ref{the longitudinal the secon order}) and
eq.(\ref{the transverse the secon order}), the first term in each
equation is gauge invariant because it is comprised of $\bar
A^{aT}_{i,q}(r)$ only.

One can choose the integration constant $f^a_q$ as
\begin{equation}
f^a_q=\frac{i}{2}g\epsilon^{abc}\frac{q_j}{q^2}\int^\infty_{-\infty}
d^3p p_j\bar \phi^b_p \bar \phi^c_{q-p} + f^{a\prime}_q
\end{equation}
with another arbitrary function $f^{\prime}_q$ and it can absorbed
into $\tilde \phi^a_{q(H)}$ by a redefinition
\begin{equation}
\tilde \phi^a_{q(H)} \rightarrow \tilde \phi^a_{q(H)} -if^{a\prime}_q.
\end{equation}

At this point it is worth pointing out that under such choice, gauge
parameters dependent parts of the total solution(\ref{total
  solutions}) has exactly the same form as the gauge
transformation(\ref{GAUGE TRANSFORM}) except the fact that the gauge
parameters are $r$ independent(See
eq.(\ref{first-order-gauge-transform}) for the gauge transformation in
$O(\epsilon)$ and eq.(\ref{second-order-gauge-transform}) in
$O(\epsilon^2)$).  Since the bulk action is manifestly gauge invariant
under the residual gauge transformation, if we plug in this 
solution into the bulk action, the gauge parameter dependent
parts drop out and the bulk on-shell action is written purely in terms
of gauge invariant parts of the total solution.

It has already been noted in the past
\cite{Witten2,Leigh1,Yee1,Donald1} that imposition of the Neumann
boundary condition on the $AdS$ boundary, leads to an ambiguity in the
computation of correlation functions of the dual operators.  This
ambiguity is associated with the residual gauge symmetry surviving at
the boundary. However, we want to look at the boundary on-shell action,
and this ambiguity appears as a total derivative term in the boundary action as long as the
current coupled to the boundary value of the Yang-Mills field is covariantly conserved, $\mathcal D_i F^a_{ri}=0$. 

\section{Boundary Conditions and the Effective Action}
\label{Boundary Conditions and the Effective Action}\setcounter{equation}{0}
In the previous section, we have discussed the bulk solution in the
radial gauge $A^a_r=0$. In this section, we would like to discuss
boundary deformations due to the bulk
solutions that we obtained in the previous section.
% For the systematic discussion, we discuss the boundary conditions
% with the leading order solution first and subleadings in order.

Before we get into the detailed discussion, we briefly discuss bulk
action.  Up to equations of motion, the bulk
action(\ref{flat-space-action}) can be written as
\begin{equation}
\label{bulk action up equation of motionN}
S[A]=\frac{1}{2}\int d^4 x \left(\partial_M(A^a_N F^a_{MN})+
\frac{1}{2}g\epsilon^{abc}A^a_M F^b_{NM}A^c_N\right).
\end{equation}
We do not need to add any counter
terms\cite{Balasubramanian1,Clifford1} since there are no divergences
in the $r\to 0$ limit, which is manifest from the bulk solutions obtained in the previous section\footnote{There is another way of adding
  counterterm action subtracting all the terms in the boundary action
  at any finite $r$ slice as \cite{Dileep1} which is indeed cut-off
  independent action.}.
We are interested in studying small $r$ behaviour (equivalently
behaviour near the $AdS$ boundary). 
%% Inside the second term in the action(\ref{bulk action up
%%   equation of motionN}), there are $r$-derivatives, so the second term
%% causes ambiguities to the total derivative terms with respect to
%% $r$. 
Both terms in the action contain radial derivatives and can be written
as total derivative with respect to $r$ which would result in boundary
contribution.  However, once we choose the radial gauge, the action
becomes
\begin{equation}
\label{radial gauge action}
S[A]=\frac{1}{2}\int d^4 x \left(\partial_r(A^a_i F^a_{ri})+
\partial_j(A^a_i F^a_{ji})+\frac{1}{2}g\epsilon^{abc}A^a_i 
F^b_{ji}A^c_j\right).
\end{equation}
The second term in eq.(\ref{radial gauge action}) then
becomes independent of $r$ derivatives and the only place where
$r$-derivative appears in the first term, and that too as total
derivative.  The third term also contributes to small $r$ boundary even if it is not total derivative with respect to $r$.
In general, it is non-trivial to extract its boundary contributions out but by using our perturbative solutions, we can evaluate those upto cubic order in small
amplitude expansion(The precise expresssion will be given in Sec.\ref{Boundary Deformation in the Second Order in}). 
As a result contribution of the bulk Yang-Mills action up
to the bulk equations of motion to small $r$-boundary is given by
\begin{equation}
\label{boundary action from bulk}
S_{bulk}\equiv\frac{1}{2}\int d^3 x A^a_i(r,x) F^a_{ri}(r,x)+\frac{1}{4}\int d^3xdr g\epsilon^{abc}A^a_i 
F^b_{ji}A^c_j.
\end{equation}
{}From now on we will call eq.(\ref{boundary action from bulk}) the
bulk action, although it is a contribution of bulk theory to the
boundary action.
% We also note that the action(\ref{bulk action up equation of
% motionN}) is manifestly gauge invariant up to equation of motion.
% This means that once we plug the solution(\ref{total solutions})
% into the action(\ref{bulk action up equation of motionN}), then the
% gauge parameters, $\bar \phi^a_q$ and $\tilde \phi^a_q$, dependent
% parts disappear.  Therefore, the action(\ref{bulk action up equation
% of motionN}) will be written in terms of the gauge invariant parts
% only.
We will mostly work in momentum space. Therefore, we perform a Fourier transform of bulk action(\ref{boundary action from bulk}) using eq.(\ref{Fourier Transform}) and we define a new
bulk action as
\begin{equation}
\hat{S}_{bulk}\equiv\frac{S_{bulk}}{(2\pi)^3},
\end{equation}
where $S_{bulk}$ is a momentum space expression of the bulk action. We define $\hat{S}_{bulk}$ to remove $(2\pi)^3$ factor from $S_{bulk}$ and
$\hat{S}_{bulk}$ will be mostly used for the construction of boundary action .
%For further discussion, we use $\hat{S}_{bulk}$.

% Since we have hired the radial gauge, then the field strength
% $F^a_{ri}=\partial_r A^a_{i}$.  One can expand the

% Near $AdS_4$ boundary, in the limit that $r \rightarrow 0$, we expand Yang-Mills
% fields as
% \begin{equation}
% \label{boundary expansion of Yang-Mills field in AdS boundary}
%  A^a_i(r,x)= A^{a(0)}_i(x)+r A^{a(1)}_i(x)+O(r^2).
% \end{equation}
% Substituting this expansion in eq.(\ref{boundary action from bulk}) we get
% \begin{equation}
% S_{bulk}=-\frac{1}{2}\int d^3 x A^{a(0)}_i(x) A^{a(1)}_i(x)+O(r).
% \end{equation}

% \footnote{There are also harmless divergent terms that can be
% added\cite{Dileep1}.  We will discuss this in Sec.\ref{Harmless
% Counter terms and Cut-off independent Boundary Action}}
% The momentum space expression of this action is obtained by the Fourier
% transform using eq.(\ref{Fourier Transform}) as
% \begin{equation}
% \hat{S}_{bulk}\equiv\frac{S_{bulk}}{(2\pi)^3}=-\frac{1}{2}\int d^3 
% p A^{a(0)}_{i,p} A^{a(1)}_{i,-p},
% \end{equation}
% where $\hat{S}_{bulk}$ is the momentum space representation of the
% bulk action.
%% which does not include $(2\pi)^3$. 
One
can define the boundary value of bulk canonical momentum, $\partial_r
A^{a}_{i,q}(r)$ of Yang-Mills field $A^{a(0)}_{i,q}(r)$ as
\begin{equation}
\hat \Pi^a_{i,q}\equiv\frac{\delta \hat{S}_{bulk}}{\delta A^{a(0)}_{i,q}}%=-A^{a(1)}_{i,-q}
\end{equation}
% The precise expression of the canonical momentum would be given
% under the condition that the regularity is precisely imposed.

The boundary on-shell action, $I_{os}$ can be defined by choosing
specific boundary conditions.  To fix the boundary condition, we
add the boundary action, $S_{bdy}$ to the bulk action as
\begin{equation}
\label{DEFINITION of ON-SHELL action}
I_{os}=S_{bulk}+S_{bdy},
\end{equation}
where we want $S_{bdy}$ is composed of the boundary value 
of the gauge invariant part of the total solution(\ref{total solutions}) and
that of its conjugate momentum only.
Then, the on-shell action is a functional of $A^{a}_{i}$ and its canonical
momentum $\Pi^a_{i}$.
% For the momentum space expression, we would use the notation
% $\hat{S}$ for every piece of action appearing.
After adding $S_{bdy}$, the generating functional for the boundary
$CFT$ will have two integration measures with $A^{a(0)}_{i}$ and
$\Pi^{a}_{i}$ as
\begin{equation}
Z[J]=e^{-W[J(A^{a(0)}_{i},\Pi^{a}_{i})]}=\int D[ A^{a(0)}_{i},
\Pi^{a}_{i}]exp\left( -S_{bulk}(A^{a(0)}_{i}) - S_{bdy}
( A^{a(0)}_{i}, \Pi^{a}_{i}) \right).
\end{equation}
The generating functional, $W[J]$ with a source $J$ is defined as
\begin{equation}
\label{on-shell action and the generating functionals}
W[J(A^{a}_{i},\Pi^a_{i})]\equiv I_{os}[A^{a}_{i},\Pi^a_{i}],
\end{equation}
where the source $J$ is again a non-trivial function of $A^{a(0)}_{i}$
and $\Pi^{a}_{i}$ in general.  The boundary conditions are given at
the saddle point of the on-shell action:
\begin{equation}
\label{saddle point on on-shell action}
\frac{\delta I_{os}[A^{a(0)}_{i},\Pi^a_{i}]}{\delta A^{a(0)}_{i}}=0
{\rm \ and \ }\frac{\delta I_{os}[A^{a(0)}_{i},\Pi^a_{i}]}{\delta \Pi^{a}_{i}}=0,
\end{equation}
and in terms of the generating functional, which is given by
\begin{equation}
\frac{\delta W[J(A^{a(0)}_{i},\Pi^a_{i})]}{\delta
  J[A^{a(0)}_{i},\Pi^a_{i}]}
\frac{\delta J(A^{a(0)}_{i},\Pi^a_{i})}{\delta A^{a(0)}_i}=0 {\rm\ and \ }
\frac{\delta W[J(A^{a(0)}_{i},\Pi^a_{i})]}{\delta
  J[A^{a(0)}_{i},\Pi^a_{i}]}
\frac{\delta J(A^{a(0)}_{i},\Pi^a_{i})}{\delta \Pi^{a}_i}=0.
\end{equation}
This corresponds to the vacuum states of the boundary $CFT$.
eq.(\ref{saddle point on on-shell action}) provides a relation between
$A^{a(0)}_{i}$ and $\Pi^a_{i}$.  Using this, one can re-write the
on-shell action in terms of $A^{a(0)}_{i}$ as saddle point
approximation.  
%If $\Pi^a_{i}$ appears in quadratic form in the
%on-shell action, this substitution is exact even quantum mechanically.

The boundary effective action can be obtained by
Legendre transform defined as
\begin{equation}
\label{Legendre transform}
\Gamma[\sigma]=-\int J\sigma+W[J],
\end{equation}
where $\Gamma$ is the boundary effective action and $\sigma$ is the
vacuum expectation value of certain boundary operators. From this
relation, one gets
\begin{equation}
\sigma=\frac{\delta W[J]}{\delta J} {\rm\ \ and \ \ } J=-\frac{\delta 
\Gamma[\sigma]}{\delta \sigma}.
\end{equation}
Now, let us suppose that for a certain boundary deformation, $S_{bdy}$,
the effective action changes in the following way
\begin{equation}
\label{deformed Gamma}
\tilde\Gamma[\sigma]=\Gamma[\sigma]+\int d^d x f(\sigma(x)),
\end{equation}
where $\Gamma$ denotes the effective action before the deformation and
$\tilde \Gamma$ denotes that after the deformation. $f$ is a function
of the vacuum expectation value $\sigma$.  The relation between $f$
and $S_{bdy}$ will become clear momentarily. Varying both sides of
(\ref{deformed Gamma}), one obtains the expression for the deformed
source $\tilde J \equiv -\frac{\delta \tilde\Gamma[\sigma]}{\delta
  \sigma}$ as
\begin{equation}
\tilde J=J-\frac{d f(\sigma)}{d \sigma}.
\end{equation}
Finally, the deformed generating functional $\tilde W[\sigma] = \tilde
\Gamma[\sigma]+\int \tilde J \sigma$ can be written as
\begin{equation}
\tilde W[\tilde J]=W[J]+\int d^d x \left(f(\sigma)-\sigma f(\sigma)\right).
\end{equation}
It is now clear from the definition of the on-shell
action(\ref{DEFINITION of ON-SHELL action}) and (\ref{on-shell action
  and the generating functionals}) that
\begin{equation}
  S_{bdy}=\int d^d x \left(f\left(\frac{\delta W[J]}{\delta J}\right)-
    \frac{\delta W[J]}{\delta J} f\left(\frac{\delta W[J]}{\delta J}\right)\right).
\end{equation}

In next section, we use these relations to derive $I_{os}$, $W[J]$ and
$\Gamma[\sigma]$ for the various deformations from $SU(2)$ Yang-Mills
theory in $AdS_4$.  Before going on, we note that the effective action
of $\Delta_+$ theory has the same form as the on-shell action of
$\Delta_-$ theory. In the case of $S_{bdy}=0$, the only possible
boundary condition is the Dirichlet boundary condition, which gives
us the $\Delta_+=2$ theory.  As we will see, to obtain the Neumann
boundary condition, we will have to set $S_{bdy}=-\int d^d x \Pi^a_i
A^{a(0)}_i$.  Since $\Pi^a_i$ is canonically conjugate of
$A^{a(0)}_i$, adding this boundary term results in Legendre transform
from $\Delta_+$ theory to $\Delta_-$ theory.  Imposition of the
Neumann boundary condition therefore results in the $\Delta_-=1$
theory.  Thus we have argued that Legendre transform of the generating
functional $W[J]$ gives us the classical effective action
$\Gamma[\sigma]$.  Therefore, the effective action of $\Delta_+$
theory should be the same with the on-shell action of $\Delta_-$
theory.

\subsection{Boundary Deformations in the First Order in $\varepsilon$}
As a warm up, we start with bulk solutions with truncations up to
$O(\varepsilon)$ and derive their on-shell actions, generating
functionals and boundary effective actions. Since, we are considering
the non-abelian gauge theory case, we explicitly write the gauge group
indices, however, up to $O(\varepsilon)$, the precess is almost the
same with the abelian gauge theory on $AdS_4$\cite{Sebastian1}.  The
only difference is that we have 3 copies of them. Therefore, the
genuine properties of the boundary effective action from $SU(2)$
Yang-Mills on $AdS_4$ will appear from the second order in
$\varepsilon$ onwards, which would be discussed in the next subsection.

The bulk solution in the first order in $\varepsilon$ in momentum
space would be expanded near $AdS$ boundary as
\begin{equation}
\label{small-r-expansion-in-momentum-space}
A^{a}_{i,q}=A^{a(0)}_{i,q}+r A^{a(1)}_{i,q}+O(r^2),
\end{equation}
where
\begin{equation}
\label{boundary expansion supple}
A^{a(0)}_{i,q}=\varepsilon \bar A^{aT(0)}_{i,q} {\rm \ and \ }
A^{a(1)}_{i,q}=\varepsilon \bar A^{a(1)}_{i,q}=-\varepsilon|q| 
\bar A^{aT(0)}_{i,q}.
\end{equation}
In eq.(\ref{boundary expansion supple}), we have used the regularity
condition(\ref{regularity-condition}) for the last equality.  As
discussed in the last section, we only deal with the gauge invariant
parts of the solutions.  The bulk action up to the bulk EOM is given
by
\begin{equation}
\label{D and N on-shell}
\hat S_{bulk}=\frac{1}{2}\varepsilon^2\int d^3 p \bar A^{a(0)}_{i,p} 
\bar A^{a(1)}_{i,-p} =-\frac{1}{2}\varepsilon^2\int d^3 p |p| 
\bar A^{a(0)}_{i,p} \bar A^{a(0)}_{i,-p}.
\end{equation}
With this expression, one can find the canonical momentum of
the boundary Yang-Mills field $A^{a(0)}_{i,q}$, which is given by
\begin{equation}
\hat \Pi^a_{i,q}=\frac{\delta \hat{S}_{bulk}}{\delta A^{a(0)}_{i,q}}
=\frac{\delta \hat{S}_{bulk}}{\delta \varepsilon \bar A^{a(0)}_{i,q}}
=-\varepsilon |q|\bar A^{aT(0)}_{i,-q}= -|q|A^{a(0)}_{i,-q}=A^{a(1)}_{i,-q}.
\end{equation}
Variation of the bulk action with respect to the boundary
field $A^{a(0)}_{i,q}$ is then given by
\begin{equation}
\delta \hat S_{bulk} = \int d^3 p  |p| \delta  A^{a(0)}_{i,p} 
A^{a(0)}_{i,-p}= \int d^3 p  \delta  A^{a(0)}_{i,p} \hat \Pi^a_{i,p}.
\end{equation}

{\em Dirichlet\ and\ Neumann\ Boundary\ Conditions}: For the case that
$\hat S_{bdy}=0$, a possible boundary condition is the Dirichlet boundary
condition, $\delta A^{a(0)}_{i,q}=0$. In this case, the on-shell
action(also the generating functional) is the same as $\hat
S_{bulk}$ and the source $J$ and the corresponding vacuum expectation
value, $\sigma$ in the generating functional are
\begin{equation}
J_D= A^{a(0)}_{i,q}  {\rm \ \ and \ \ }\sigma_D \equiv\frac{\delta W[J_D]}{\delta J_D}
=\frac{\delta \hat S_{bulk}}{\delta A^{a(0)}_{i,q}}=\hat \Pi^a_{i,q}=-|q| A^{a(0)}_{i,-q},
\end{equation}
respectively, where the subscript $D$ denotes ``Dirichlet''. The
boundary effective action can be obtained by Legendre transform
defined in eq.(\ref{Legendre transform}).
% \begin{equation}
% \label{legendre transform}
% \Gamma[\sigma]=-\int J\sigma+W[J],
% \end{equation}
% where $\Gamma$ becomes the boundary effective action in terms of the
% vacuum expectation value $\sigma$.  
We apply the Legendre transform
for the Dirichlet case, then the effective action is given by
\begin{equation}
\label{Dirichlet boundary effective order varepsilon}
\Gamma^D[\hat \Pi^a_{i,q}]=\frac{1}{2} \int^\infty_{-\infty}
\frac{d^3 p}{|p|}\hat\Pi^a_{i,p}\hat\Pi^a_{i,-p}.
\end{equation}

Neumann boundary condition can be obtained by considering that $\hat
S^N_{bdy}=-\int d^3 p A^{a(0)}_{i,p} \hat\Pi^a_{i,p}$, where the
superscript $N$ denotes ``Neumann''.  To find out stationary points,
we vary $I^N_{os}[ A^{a(0)}_{i,q},\Pi^a_{i,q}]$ as
\begin{equation}
\delta I^N_{os}[A^{a(0)}_{i,q},\hat\Pi^a_{i,q}] = \int d^3 p  
\delta  A^{a(0)}_{i,p} \Pi^a_{i,p} + \delta S_{bdy}
= -\int d^3 p   A^{a(0)}_{i,p} \delta \hat\Pi^a_{i,p}=0,
\end{equation}
so we get Neumann boundary condition: $\delta \hat\Pi^a_{i,q}=0$. For
Neumann case, the role of the source $J$ and the vacuum expectation
value $\sigma$ are interchanged with respect to the Dirichlet case. This is because
adding $\hat S^N_{bdy} = \int J\sigma$ is effectively performing
Legendre transform of $\hat S_{bulk}$.  As a result, the
boundary effective action is obtained from Legendre transformation of
eq.(\ref{Dirichlet boundary effective order varepsilon}):
\begin{equation}
\Gamma^N[A^{a(0)}_{i,p}]=-\frac{1}{2}\int d^3 p |p| A^{a(0)}_{i,p} 
A^{a(0)}_{i,-p},{\ \ }J_N=\hat\Pi^a_{i,q} 
{\rm \ \ and\ \ }\sigma_N=A^{a(0)}_{i,p}.
\end{equation}

{\em Massive\ Deformation}: One can also discuss generalized Neumann
boundary conditions, for example, the {\em Massive\ Deformation}. 
At the first order in $\varepsilon$, the massive deformation leads to
a boundary condition given by
\begin{equation}
\label{Massive deformation}
\bar A^{a(0)T}_{ip}+\frac{1}{m}\bar A^{a(1)}_{ip}=0.
\end{equation}
To obtain this boundary condition, we introduce the boundary action
\begin{equation}
\hat S^M_{bdy}=-\int d^3 p  \left( A^{a(0)}_{i,p} \hat\Pi^a_{i,p}+
\frac{1}{2m}\hat\Pi^a_{i,p}\hat\Pi^a_{i,-p}\right).
\end{equation}
By varying the on-shell action with above boundary action, we end up
with
\begin{equation}
\label{delta I^N_os}
\delta I^M_{os}[A^{a(0)}_{i,q},\Pi^a_{i,q}]=-\int d^3 p 
\delta A^{a(0)}_{i,p}\left( |p| A^{a(0)T}_{i,-p}+\hat\Pi^a_{i,p}\right)
-\int d^3 p \delta \Pi^a_{i,-p}\left( A^{aT(0)}_{i,-p}+
\frac{1}{m}\Pi^a_{i,p} \right)=0,
\end{equation}
where the first integral gives the regularity
condition.
%vanishes with definition of the canonical momentum $\Pi^a_{i,q}$.
Rather than imposing Neumann boundary condition for the second
integration, if we set the quantity inside the parenthesis to zero,
then the canonical momentum becomes
\begin{equation}
\label{massive canonical momentum constraint}
\hat\Pi^a_{i,-q}=-m A^{a(0)T}_{i,q}.
\end{equation}
% Once this boundary condition is imposed, then the boundary effective
% action becomes the on-shell action.
For the consistency with the regularity
condition (\ref{regularity-condition}), it is demanded that $|p|=m$.
Therefore, the boundary field $A^{a(0)}_{i,q}$ becomes on-shell and
massive under such a condition.

We rewrite the on-shell action $I^M_{os}$ with replacing every
$\Pi^a_{i,q}$ by $A^{a(0)}_{i,q}$ using eq.(\ref{massive canonical
  momentum constraint})
% \footnote{In this substitution, we only use Eq.(\ref{massive
% canonical momentum constraint}) and do not use the regularity
% condition. This is because the regularity condition becomes one of
% the boundary conditions imposed in Eq.(\ref{delta I^N_os}). What we
% want to is to evaluate the on-shell action in terms of
% $A^{a(0)}_{i,q}$ by imposing only one of the two boundary conditions
% in Eq.(\ref{delta I^N_os}).  Then, one can impose the regularity
% condition only as boundary condition or does Eq.(\ref{massive
% canonical momentum constraint}) without regularity condition.},
as
\begin{equation}
\label{massive effective action in order epsilon}
I^M_{os}[A^{a(0)}_{i,q}]=-\frac{1}{2}\int d^3 p 
\left( |p|-m \right)A^{a(0)T}_{i,p}A^{a(0)T}_{i,-p}.
\end{equation}
% This substitution makes sense even at the quantum level, because in
% the on-shell action, $\Pi^{a,(0)}_{i,q}$ appears quadratically, so one
% can perform the Gaussian integration.  
The fact that this procedure is
justified can be seen by varying $I^M_{os}[A^{a(0)}_{i,q}]$ with
respect to $A^{a(0)}_{i,q}$ and noticing that it produces the correct
boundary condition
\begin{equation}
\frac{\delta I^M_{os}[A^{a(0)}_{i,q}]}{\delta A^{a(0)}_{i,q}}=-
\left( |p|-m \right)A^{a(0)}_{i,-p}=0.
\end{equation}
% This also gives the correction boundary condition, $|p|=m$ and
% actually under this condition $I^M_{os}=W^M[J(A^{a(0)}_{i,q})]=0$.
The final step for the massive deformation case is to obtain the dual
$CFT$ (or effective) action.
% To demonstrate how to got the dual $CFT$ action, we keep the form of
% Eq.(\ref{massive effective action in order epsilon}) without setting
% $|p|=m$ in it.
Unfortunately, one cannot easily figure out what is the deformed
source $J$ in above expression and therefore cannot perform Legendre
transform either.  However, there is another way to deal with this
situation where one writes down an expected form of the dual $CFT$
action. Let us consider the following form:
\begin{equation}
\Gamma^M[A^{a(0)}_{i,q}]=\frac{1}{2}\int d^3 p 
\alpha(p)A^{a(0)}_{i,p}A^{a(0)}_{i,-p},
\end{equation}
where $\alpha$ is an arbitrary momentum dependent function and we
assume that vacuum expectation value $\sigma$ is still
$A^{a(0)}_{i,q}$ under any deformation\cite{Ioannis1}.  Using this for
the effective action, one can derive the expression of the source 
\begin{equation}
J_M[A^{a(0)}_{i,q}]=-\frac{\delta \Gamma^M[J(A^{a(0)}_{i,q})]}{\delta 
A^{a(0)}_{i,q}}=-\alpha(q)A^{a(0)}_{i,-q}.
\end{equation}
We can then use this source term $J$ to perform inverse Legendre
transform from $\Gamma$ to obtain the generating functional $W$
using eq.(\ref{Legendre transform}).  We then demand that this inverse
transformation reproduce the correct generating functional $W$, which
imposes a constraint on $\alpha$, and also determines expression of
the source term,
\begin{equation}
\label{alp-sou}
\alpha=(|p|-m){\ , \ }J_M=-(|p|-m)A^{a(0)}_{i,-q}{\rm\ and\ }
\Gamma^M=\frac{1}{2}\int d^3 p (|p|-m)A^{a(0)}_{i,p}A^{a(0)}_{i,-p}.
\end{equation}
The generating functional $W$ is usually expressed as the functional
of source $J_M$, which is done by using eq.(\ref{alp-sou}),
\begin{equation}
 W^M[J^M_{i,q}]=\frac{1}{2}\int d^3 p \frac{J^{M}_{i,p}J^{M}_{i,-p}}{ |p|-m}.
\end{equation}

{\em Self-Dual\ Boundary\ Condition\ and\ Massive\ Deformation}: The
most interesting case is the self-dual boundary condition, together
with the massive deformation.  Self-duality condition in four dimension is
given by
\begin{equation}
\label{self-dual-condition}
F^a_{MN}=\frac{1}{2}\epsilon_{MNPQ}F^a_{PQ}.
\end{equation}
To study self-dual boundary condition, we expand Yang-Mills
field near the $AdS$ boundary, {\em i.e.}, around $r=0$ as in
eq.(\ref{small-r-expansion-in-momentum-space}).
% \begin{equation}
% A^a_{M}=A^{a(0)}_{M}+rA^{a(1)}_{M}+O(r^2).
% \end{equation}
Once we choose the index $M=r$ in eq.(\ref{self-dual-condition}), the
boundary condition derived from it becomes
\begin{equation}
\label{self-dual-in-boundary}
A^{a(1)}_i=\mathcal D_iA^{a(0)}_r+\frac{1}{2}\epsilon_{ijk}F^{a(0)}_{jk},
\end{equation}
where $\mathcal
D_iA^{a}_r=\partial_iA^{a}_r-g\epsilon^{abc}A^{b}_iA^{c}_r$.
% (We also need to check if our solution can satisfy this self-duality
% condition.It probably satisfies.)
Since we have used the radial gauge $A^a_r=0$ for our bulk solutions,
$\mathcal D_iA^{a}_r=0$ in eq.(\ref{self-dual-in-boundary}).
% Under the radial gauge, in momentum space,
% Eq.(\ref{self-dual-in-boundary}) is written as
% \begin{equation}
% A^{a(1)}_{i,q}=\frac{1}{2}\epsilon_{ijk}\left( -iq_j A^{a(0)}_{k,q}
% +iq_k A^{a(0)}_{j,q}-g\epsilon^{abc}\int^\infty_{-\infty}d^3 
% p A^{b(0)}_{j,p}A^{c(0)}_{k,q-p}\right),
% \end{equation}
% Next we discuss constraints on the boundary from the equation of
% motion. In Eq.(\ref{equation-of-motion}), we set $N=r$, then the
% leading order EOM in the boundary expansion becomes
% \begin{equation}
% \label{boundary-constraint}
% 0=\partial_i F^{a(0)}_{ir}+g\epsilon^{abc}F^{b(0)}_{ir}A^{c(0)}_{i},
% \end{equation}
% where
% $F^{a(0)}_{ir}=\partial_iA^{a(0)}_{r}-A^{a(1)}_{i}-g
% \epsilon^{abc}A^{b(0)}_iA^{c(0)}_{r}$
% and we define that $E^{a}_i$ is boundary electric field as
% $E^{a}_{i}\equiv F^{a}_{ri}$. Plug Eq.(\ref{boundary-constraint})
% into Eq.(\ref{self-dual-in-boundary}), then we obtain the boundary
% self-duality as
% \begin{equation}
% \frac{1}{2}\epsilon_{ijk}F^{a(0)}_{jk}=F^{a(0)}_{ri},{\rm 
% \ similarily\ \ }B^a_i=E^a_i,
% \end{equation}
% where $B_{i}\equiv \frac{1}{2}\epsilon_{ijk}F^{a(0)}_{jk}$. All
% these fields would be interpreted as fields in dual $CFT$.
%
% For further discussion, we would like to use radial gauge,
% $A^a_{r}=0$.
Up to the leading order in $\varepsilon$, the self dual
boundary condition is given by
\begin{equation}
\label{Self=duality}
A^{a(1)}_i(x)= \epsilon_{ijk}\partial_j A^{a(0)}_{k}(x), {\rm \ in 
\ momentum\ space\ } 
A^{a(1)}_{i,q}=\hat \Pi^a_{i,-q}= \epsilon_{ijk}(-iq_j)A^{a(0)}_{k,q}.
\end{equation}
In addition to this, if we impose the on-shell condition,
$(|p|-m)A^{a(0)}_{i,p}=0$, it gives rise to massive deformation of the
boundary on-shell action.  That is, eq.(\ref{Massive deformation})
together with eq.(\ref{Self=duality}), gives rise to the boundary
condition 
\begin{equation}
\label{massive-sef-duality-CONdITion}
0=mA^{a(0)}_{i,p}+\epsilon_{ijk}(-ip_j)A^{a(0)}_{k,p}.
\end{equation}

This boundary condition can be incorporated in boundary on-shell
action in the following way,
\begin{equation}
\hat S^{MS}_{bdy}=\int d^3 p \left[ \beta \left( A^{a(0)}_{i,p} \hat\Pi^a_{i,p}
+\frac{1}{2m}\hat\Pi^a_{i,p}\hat\Pi^a_{i,-p} \right)
-\frac{\beta+1}{2}\epsilon_{ijk}A^{a(0)}_{i,p}(ip_j)A^{a(0)}_{k,-p}\right],
\end{equation}
where $\beta$ is a numerical parameter. Variation of the on-shell
action, $I^{MS}_{os}=\hat S_{bulk}+\hat S^{MS}_{bdy}$, provides
\begin{eqnarray}
\label{sd-massive}
\delta I^{MS}_{os}[A^{a(0)}_{i,q},\hat\Pi^a_{i,q}]&=& \int d^3 p 
\beta\delta \hat\Pi^a_{i,p}\left( A^{a(0)}_{i,p}+\frac{1}{m}
\hat\Pi^a_{i,-p} \right) \\ \nonumber
&-& \int d^3 p \delta A^{a(0)}_{i,p}\left( |p|A^{a(0)}_{i,-p}-
\beta\hat\Pi^a_{i,p}+(\beta+1)\epsilon_{ijk}(ip_j)A^{a(0)}_{k,-p}\right).
\end{eqnarray}
The first line in above equation (\ref{sd-massive}) can be set to zero
by considering the massive deformation
\begin{equation}
 A^{a(0)}_{i,q}+\frac{1}{m}\hat\Pi^a_{i,-q} =0
\end{equation}
rather than imposing the Neumann boundary condition, $\delta
\Pi^a_{i,q}=0$. For consistency with the regularity condition, we
demand $(|p|-m)A^{a(0)}_{i,p}=0$ and the massive deformation implies
the canonical momentum is given by $\Pi^a_{i,-q}=-m A^{a(0)}_{i,q}$.
For the second line in equation (\ref{sd-massive}), rather than
imposing Dirichlet boundary condition $ \delta A^{a(0)}_{i,q}=0$, we
equate the expression inside the parenthesis to zero.  This choice
corresponds to the self dual boundary condition in momentum
space,
\begin{equation}
\hat\Pi^a_{i,q}=-|q|A^{a(0)}_{i,-q}=\epsilon_{ijk}(-iq_j)A^{a(0)}_{k,-q}.
\end{equation}
This condition together with on-shell condition, is exactly the same
with eq.(\ref{massive-sef-duality-CONdITion}).
% \begin{equation}
% A^{a(0)}_{i,-q}=\frac{1}{m}\epsilon_{ijk}(-iq_j)A^{a(0)}_{k,-q},
% \end{equation}
When we substitute this relation into
$I^{MS}_{os}[A^{a(0)}_{i,q},\hat\Pi^a_{i,q}]$ along with the
regularity condition
% \footnote{In this case, the regularity is not a boundary condition,
% of which case is different from massive deformation.}
we can eliminate $\hat\Pi^a_{i,q}$ by expressing it in terms of
$A^{a(0)}_{i,-q}$ to get
\begin{equation}
I^{MS}_{os}[A^{a(0)}_{i,q}]=-\frac{1}{2}(1+\beta)\int d^3p
\left(
  mA^{a(0)T}_{i,p}A^{a(0)T}_{i,-p}+\epsilon_{ijk}A^{a(0)}_{i,p}(ip_j)
A^{a(0)}_{k,-p}\right),
\end{equation}
which is abelian massive Chern-Simons
action\cite{Townsend1,Deser1}.  We also obtain the deformed source
and dual $CFT$ action by the same method in the previous
discussion with massive deformation. They are given by
\begin{eqnarray}
 J_{MS}&=&-(1+\beta)\left( mA^{a(0)}_{i,-q}+\epsilon_{ijk}(iq_j)
A^{a(0)}_{k,-q}\right), \\
\Gamma^{MS}[A^{a(0)}_{i,q}]&=&\frac{1}{2}(1+\beta)\int d^3p\left( 
mA^{a(0)T}_{i,p}A^{a(0)T}_{i,-p}+\epsilon_{ijk}A^{a(0)}_{i,p}(ip_j)A^{a(0)}_{k,-p}\right).
\end{eqnarray}

\subsection{Boundary Deformation in the Second Order in $\varepsilon$}
\label{Boundary Deformation in the Second Order in}
A way of imposing boundary conditions for the second order solution in
$\varepsilon$ is in principle the same with previous discussion.
There are some technical difficulties due to appearance of quadratic
terms involving the first order solutions.  However, this nonlinearity
in the equation involves lower order solutions only, which are already
derived using the small amplitude expansion.  
For evaluating the boundary on-shell action, we would like to choose a gauge for boundary gauge fields, 
$A^{a(0)}_{i,q}$, in fact, we will
set $\phi^a_{i,q}=\varepsilon \bar \phi^a_{i,q} +\varepsilon^2 \tilde \phi^a_{i,q}=0$. 
Since the bulk action is manifestly gauge invaraint, choosing a particular gauge is not a problem. 
With such choice of gauge degree of freedom, the boundary gauge field appearing on the boundary on-shell action will be effectively transverse.
Therefore, in the following, we only deal with gauge parameter independent 
parts of the solutions for the construction of the boundary theory.
% \footnote{Once we plug the bulk solutions into (\ref{radial gauge action}), 
% gauge parameter dependent terms will vanish since the bulk action is manifestly invariant under residual gauge transform. 
% The boundary contributions from the bulk action would be also comprised only of gauge parameter independent parts of the bulk solutions. 
% However, bulk action (\ref{boundary action from bulk}) is not manifestly gauge invariant even when one imposes conservation of the dual current of the boundary source term, 
% $\mathcal D_i F^a_{ri}=0$. 
% In fact, to see gauge invariance, we need to keep all the bulk terms as (\ref{radial gauge action}) together with bulk equations of motion.}. 
We start with a general
discussion of the solution(\ref{total solutions}).  The near $AdS$
boundary expansion is given by
% \begin{eqnarray}
% A^{a}_{i,q}(r)|_{r\rightarrow 0}&=&A^{a(0)}_{i,q}+rA^{a(1)}_{i,q}+O(r^2)
%  \\  \nonumber
% &=&A^{a(0)}_{i,q}-|q|rA^{a(0)}_{i,q}-rg\epsilon^{abc}
% \int^{\infty}_{-\infty}d^3 p A^{c(0)}_{k,p} A^{b(0)}_{j,q-p}\Delta_{ijk}(p,q)
% %\left( \frac{\alpha_{ijk}(p,q)}{|p|+|q-p|+|q|}\right. \\ \nonumber
% %&-&\left.\frac{q_i\delta_{jk}|q-p|(|q-p|+|p|-|q|)}{q^2(|p|+|q-p|)}\right) + O(r^2),
% \end{eqnarray}
\begin{equation}
A^{a}_{i,q}(r)|_{r\rightarrow 0}=A^{a(0)}_{i,q}+rA^{a(1)}_{i,q}+O(r^2),
\end{equation}
where
\begin{equation}
A^{a(1)}_{i,-q}=
-|q|A^{a(0)}_{i,-q}-g\epsilon^{abc}\int^{\infty}_{-\infty}d^3 p 
A^{c(0)}_{k,p} A^{b(0)}_{j,-q-p}\Delta_{ijk}(p,-q)+O(\varepsilon^3)
\end{equation}
and
\begin{equation}
\label{definition of Delta}
\Delta_{ijk}(p,q)=\frac{\alpha_{ijk}(p,q)}{|p|+|q-p|+|q|}-\frac{iq_i
\delta_{jk}|q-p|(|q-p|+|p|-|q|)}{q^2(|p|+|q-p|)}.
\end{equation}
$A^{a(0)}_{i,q}$ is the boundary value of the full solution
$A^{a}_{i,p}(r)$ defined in eq.(\ref{total solutions})(See also
eq.(\ref{the transverse the secon order}), eq.(\ref{the longitudinal the
  secon order}) and eq.(\ref{the second order homogeneous solution})),
which is given by
\begin{eqnarray}
\label{A0exp}
A^{a(0)}_{i,q}&=&\varepsilon \bar A^{aT(0)}_{i,p}
+ \varepsilon^2 \tilde A^{aT(0)}_{i,p(H)}
-\varepsilon^2
ig\epsilon^{abc}\frac{q_i}{q^2}\int^\infty_{-\infty}d^3p 
\bar A^{cT(0)}_{j,p} \bar A^{bT(0)}_{j,q-p}\frac{|q-p|}{|p|+|q-p|} \\ \nonumber
&+&\varepsilon^2 g\epsilon^{abc}\int^\infty_{-\infty}d^3p 
\bar A^{cT(0)}_{k,p} \bar A^{bT(0)}_{j,q-p}\frac{\alpha_{ijk}(p,q)}
{(|p|+|q-p|)^2-q^2}+O(\varepsilon^3).
\end{eqnarray}
Now, we evaluate the bulk action(\ref{boundary action from bulk})
explicitly by substituting the bulk solution and keeping terms upto
the leading interaction terms, 
\begin{equation}
\label{bulk-second-sol}
S_{bulk}\equiv\frac{1}{2}\int d^3 q A^{a(0)}_{i,q} A^{a(1)}_{i,-q}+\frac{1}{4}\int d^3q d^3l d^3 p dr g\epsilon^{abc}A^a_{i,q} 
F^b_{ji,l}A^c_{j,p}\delta^3(q+l+p),
\end{equation}
where gauge fields in the second term contains the first order
solutions only, which means that 
\begin{equation}
A^a_i=\varepsilon \bar A^{a(0)}_{i,q}e^{-|q|r} + O(\varepsilon^2)=A^{a(0)}_{i,q}e^{-|q|r}+ O(\varepsilon^2).
\end{equation}
Therefore, the second term in eq.(\ref{bulk-second-sol}) becomes
\begin{eqnarray}
S^{2nd \ term}_{bulk}\!\!\!\!\!
% &=&\frac{1}{2}\int drd^3q d^3l d^3 p  g\epsilon^{abc}\varepsilon^3 \bar A^{a(0)}_{i,q} 
% (-il_j\bar A^{b(0)}_{i,l})\bar A^{c(0)}_{j,p}e^{-(|q|+|l|+|p|)r}\delta^3(q+l+p)\\ \nonumber
% &+&O(\varepsilon^4)\\ \nonumber
&=&\frac{1}{2}\int d^3q d^3l d^3 p \left.\varepsilon^3 \bar A^{a(0)T}_{i,q} 
\bar A^{b(0)T}_{j,l}\bar A^{c(0)T}_{k,p}\frac{il_k \delta_{ij}}{|q|+|l|+|p|}e^{-(|q|+|l|+|p|)r}\delta^3(q+l+p)\right|^{r=0}_{r=\infty}\\ \nonumber
&=&\frac{1}{2}\int d^3q d^3l d^3 p A^{a(0)}_{i,q} 
 A^{b(0)}_{j,l} A^{c(0)}_{k,p}\frac{il_k \delta_{ij}}{|q|+|l|+|p|}\delta^3(q+l+p)+O((A^{a(0)}_{i,q})^4)
\end{eqnarray}

With this, one can construct the bulk action as
\begin{eqnarray}
\label{Dirichlet action}
\hat S_{bulk}&=&-\frac{1}{2} \int^\infty_{-\infty} d^3 q
|q|A^{a(0)}_{i,q} A^{a(0)}_{i,-q}
-\frac{1}{2}g\epsilon^{abc}\int^\infty_{-\infty} d^3 q d^3 p 
A^{a(0)}_{i,q} A^{b(0)}_{j,-q-p} A^{c(0)}_{k,p}\Delta_{ijk}(p,-q) \\ \nonumber
&+&\frac{1}{2}\int d^3q d^3l d^3 p A^{a(0)}_{i,q} 
 A^{b(0)}_{j,l} A^{c(0)}_{k,p}\frac{il_k \delta_{ij}}{|q|+|l|+|p|}\delta^3(q+l+p),
\end{eqnarray}
upto cubic interactions.
Notice that $\Delta_{ijk}$ and $\frac{il_k \delta_{ij}}{|q|+|l|+|p|}$, in order to be non-vanishing, should be
fully anti-symmetric in indices, $i$, $j$ and $k$ together with
appropriate momentum exchange due to $\epsilon^{abc}$. The second
term is then written as
\begin{equation}
\label{the second term}
\hat S_{2nd\ term} = -\frac{1}{2}g \epsilon^{abc}
\int^{\infty}_{-\infty}d^3 q d^3 p d^3 l A^{a(0)}_{i,q} A^{b(0)}_{j,l}
A^{c(0)}_{k,p}\delta^3(q+p+l)\tilde\Delta_{ijk}(q,l,p),
\end{equation}
where
\begin{equation}
\tilde\Delta_{ijk}(q,l,p)=\tilde\Delta^T_{ijk}(q,l,p)+
\tilde\Delta^L_{ijk}(q,l,p).
\end{equation}
$\tilde\Delta^T_{ijk}(q,l,p)$ and $\tilde\Delta^L_{ijk}(q,l,p)$ are
given by
\begin{equation}
\label{DeltaT}
\tilde\Delta^T_{ijk}(q,l,p)=\frac{i(l-q)_k \delta_{ij}+i(p-l)_i
  \delta_{jk}+i(q-p)_j \delta_{ik}}{2(|q|+|l|+|p|)}.
\end{equation}
\begin{eqnarray}
\label{DeltaL}
\tilde\Delta^L_{ijk}(q,l,p)&=&\frac{iq_i\delta_{jk}(|l|-|p|)
(|p|+|l|-|q|)}{6q^2(|p|+|l|)}
+\frac{il_j\delta_{ki}(|p|-|q|)(|q|+|p|-|l|)}{6l^2(|q|+|p|)}\\ \nonumber
&+&\frac{ip_k\delta_{ij}(|q|-|l|)(|l|+|q|-|p|)}{6p^2(|l|+|q|)},
\end{eqnarray}
(eq.(\ref{DeltaT}) and eq.(\ref{DeltaL}) can be obtained from
eq.(\ref{definition of alpha}) and eq.(\ref{definition of Delta})
after some computation using the fact that $\bar A^{aT(0)}_{i,q}$ is
transverse). In fact, $\tilde\Delta^L_{ijk}(q,l,p)$ does not
contribute to the bulk action, since the fields multiplying it in the
action are effectively transverse\footnote{The bulk solution of
  Yang-Mills fields up to second order in $\varepsilon$, requires
  terms only up to cubic in $\varepsilon$ in $\hat S_{bulk}$.  Using
  the expansion (\ref{A0exp}), of the boundary value of the Yang-Mills
  field $A^{a(0)}_{i,q}$ in the cubic interaction in
  eq.(\ref{Dirichlet action}), it is easy to see that up to
  $O(\varepsilon^3)$ this term is effectively transverse
\begin{equation}
A^{a(0)}_{i,q} A^{b(0)}_{j,-q-p} A^{c(0)}_{k,p} = \varepsilon^3 \bar
A^{aT(0)}_{i,q} \bar A^{bT(0)}_{j,-q-p} \bar A^{cT(0)}_{k,p}+O(\varepsilon^4).
\end{equation}
As a result, at this order, $\tilde \Delta^L_{ijk}(q,l,p)$ disappears
from the boundary on-shell action.}.
% One can recognize that the expression is fully anti-symmetric in
% changing boundary space-time indices contracting with $\Delta_{ijk}$
% However, each term in $\Delta_{ijk}$ contains Kronecker $\delta$
% (See Eq.(\ref{definition of alpha}) and Eq.(\ref{definition of
% Delta})), so it vanishes.  Then, it is obtained
% \begin{equation}
% \hat S_{bulk}=-\frac{1}{2} \int^\infty_{-\infty} d^3 q |q|A^{a(0)}_{i,q} A^{a(0)}_{i,-q}.
% \end{equation}
The third term in eq.(\ref{Dirichlet action}) is given by
\begin{equation}
\hat S^{3rd\ term}=\frac{1}{6}\int d^3q d^3l d^3 p A^{a(0)}_{i,q} 
 A^{b(0)}_{j,l} A^{c(0)}_{k,p}\tilde \Delta^T_{ijk}(q,l,p)\delta^3(q+l+p),
\end{equation}
it also has the same anti-symmetrization.

The canonical momentum of the source $A^a_{i,q}$ is given by
\begin{equation}
\label{the second order momentum definietion}
\hat\Pi^a_{i,q}=\frac{\delta \hat{S}_{bulk}}{\delta A^{a(0)}_{i,q}}
=-|q|A^{a(0)}_{i,-q}-g\epsilon^{abc}\int^{\infty}_{-\infty}
d^3 p A^{b(0)}_{j,-q-p}A^{c(0)}_{k,p}\tilde\Delta_{ijk}(q,-q-p,p)
\end{equation}

{\em Dirichlet\ Boundary\ Condition}: Without adding any boundary
action, the on-shell action, $I^D_{os}$ is given by
%Eq.(\ref{Dirichlet action}). 
\begin{equation}
I^{D}_{os}(A^{a(0)}_{i,q})=
-\frac{1}{2} \int^\infty_{-\infty} d^3 q |q|A^{a(0)}_{i,q} A^{a(0)}_{i,-q}
-\frac{1}{3}g\epsilon^{abc}\int^\infty_{-\infty} d^3 q d^3 p 
A^{a(0)}_{i,q} A^{b(0)}_{j,-q-p} A^{c(0)}_{k,p}
\tilde\Delta^T_{ijk}(q,-q-p,p).
\end{equation}
The Legendre transform of $I^D_{os}$ becomes the boundary effective action
in terms of dual operator $\Pi^{a}_{i,q}$, which is given by
\begin{equation}
\label{Dirichlet-epsilon-square-action}
\Gamma^{D}(\Pi^a_{i,q})=\frac{1}{2}\int^{\infty}_{-\infty}
\frac{d^3 q}{|q|}\hat\Pi^a_{i,q}\hat\Pi^a_{i,-q}
+\frac{1}{3}g\epsilon^{abc}\int^{\infty}_{-\infty}d^3 q d^3 p
\frac{\tilde \Delta_{ijk}(q,-q-p,p)}{|q||q+p||p|}
\hat\Pi^a_{i,q}\hat\Pi^b_{j,-q-p}\hat\Pi^c_{k,p}.
\end{equation}
This action has exotic momentum dependent cubic interaction,
which is classically marginal.  Up to this order, we can evaluate
2-point and 3-point functions of the boundary $CFT$ and the dual
$CFT$.
% In further discussions with different boundary conditions, we have
% obtained different types of 2-pt and 3-pt functions.

{\em Neumann\ Boundary\ Condition}: The effective action in Neumann
boundary condition can be obtained by Legendre transform
of (\ref{Dirichlet-epsilon-square-action}), which becomes
\begin{equation}
\Gamma^{N}(A^{a(0)}_{i,q})=
-\frac{1}{2} \int^\infty_{-\infty} d^3 q |q|A^{a(0)}_{i,q} A^{a(0)}_{i,-q}
-\frac{1}{3}g\epsilon^{abc}\int^\infty_{-\infty} d^3 q d^3 p 
A^{a(0)}_{i,q} A^{b(0)}_{j,-q-p} A^{c(0)}_{k,p}\tilde\Delta_{ijk}(q,-q-p,p),
\end{equation}
with $J^N=\hat\Pi^a_{i,q}$.  The generating functionals for each
boundary condition are given by
\begin{eqnarray}
I^{D}_{os}(A^{a(0)}_{i,q})=W^D[A^{a(0)}_{i,q}]=\Gamma^{N}[A^{a(0)}_{i,q}] 
{\rm \ \ and \ \ } I^{N}_{os}(\Pi^{a}_{i,q})=W^{N}(\Pi^a_{i,q})=
\Gamma^{D}[\Pi^a_{i,q}].
\end{eqnarray}

{\em Massive\ and\ Self-Dual\ Boundary\ Condition}: To impose
self-dual and massive deformation as a boundary condition, we 
add the following boundary action to $\hat S_{bulk}$:
\begin{eqnarray}
\hat S_{bdy}&=&\int^\infty_{-\infty} d^3 q \left[ -\beta  
\left(A^{a(0)}_{i,q}\Pi^a_{i,q}+\frac{1}{2m}\Pi^a_{i,q}\Pi^a_{i,-q} 
\right) \right.\\ \nonumber
&+&\frac{3}{2m}\alpha \beta g\epsilon^{abc}\Pi^a_{i,q}
\int^{\infty}_{-\infty}d^3 p d^3 l A^{b(0)}_{j,l}A^{c(0)}_{k,p}
\delta^3(-q+l+p)\tilde \Delta_{ijk}(-q,l,p)
\\ \nonumber
&+& \eta\epsilon_{ijk}\left(A^{a(0)}_{i,q}(iq_j)
A^{a(0)}_{k,-q}-\frac{1}{3}g\epsilon^{abc}\int^{\infty}_{-\infty}d^3 p 
d^3 l A^{a(0)}_{i,q}A^{b(0)}_{j,l}A^{c(0)}_{k,p}\delta^3(q+l+p)\right)\\ \nonumber
&+&\left.\frac{\gamma}{3} g\epsilon^{abc}\int^{\infty}_{-\infty}d^3p 
d^3l A^{a(0)}_{i,q}A^{b(0)}_{j,l}A^{c(0)}_{k,p}\delta^3(q+l+p)
\tilde \Delta_{ijk}(q,l,p) \right],
\end{eqnarray}
where $\alpha$, $\beta$, $\gamma$ and $\eta$ are numerical constants which
would be determined by imposing right boundary condition.  Variation
of $I^{MS}_{os}[A^{a(0)}_{i,q}]=\hat S_{bulk}+\hat S_{bdy}$ with
respect to $A^{a(0)}_{i,q}$ and $\hat\Pi^a_{i,q}$ provides the
following boundary conditions:
\begin{eqnarray}
\label{massive-self-dual condition-1}
\hat\Pi^a_{i,q}&=&-mA^{a(0)}_{i,-q}+\frac{3}{2}\alpha g \epsilon^{abc}
\int^{\infty}_{-\infty}d^3 p d^3 l  A^{b(0)}_{j,l}A^{c(0)}_{k,p}
\delta^3(q+l+p)\tilde \Delta_{ijk}(q,l,p), \\
\label{massive-self-dual condition-2}
(\beta-1)\Pi^{a}_{i,q}&=&g\epsilon^{abc}\int^{\infty}_{-\infty}d^3 p 
d^3 l  A^{b(0)}_{j,l}A^{c(0)}_{k,p}\delta^3(q+l+p)
(\gamma-\frac{3\alpha\beta(|l|+|p|)}{2m})\tilde\Delta_{ijk}(q,l,p) \\ \nonumber
&+&\eta\epsilon_{ijk}F^{a}_{jk,-q},
\end{eqnarray}
where
\begin{equation}
F^{a}_{ij,q}=-iq_i A^a_{j,q}+iq_j A^a_{i,q}
-g\epsilon^{abc}\int^{\infty}_{-\infty}d^3 p d^3 l A^{b(0)}_{i,l}
A^{c(0)}_{j,p}\delta^3(-q+l+p),
\end{equation}
is Yang-Mills field strength in momentum space. For the consistency
between canonical momentum(\ref{the second order momentum
  definietion}) and boundary condition(\ref{massive-self-dual
  condition-1}), one requires
\begin{equation}
\alpha=-\frac{2}{3} {\rm \ and \ } A^{a(0)}_{iq}\rightarrow A^{a(0)}_{iq}|_{|q|=m}.
\end{equation}
% This means that the phase space of momenta of the field
% $A^{a(0)}_{i,q}$ becomes on shell, and any phase space integrations
% containing $A^{a(0)}_{i,q}$ will restricts their integration range
% under such condition. This is an extension of the massive
% deformation from abelin case\cite{Sebastian1}.
We can use the second boundary condition(\ref{massive-self-dual
  condition-2}) to impose non-abelian version of massive self-dual
boundary condition\cite{Townsend1}, which is given by
\begin{equation}
A^{a(0)}_{i,q}=-\frac{1}{2m}\epsilon_{ijk}F^{a}_{jk,q}.
\end{equation}
To do this, we plug eq.(\ref{massive-self-dual condition-1}) into
eq.(\ref{massive-self-dual condition-2}).  Then the massive self-dual
boundary condition can be obtained if we impose the condition,
\begin{equation}
 \gamma=1-3\beta {\rm\ \ and\ \ } \eta=\frac{\beta-1}{2}.
\end{equation}
% Therefore, the boundary theory that we seek is controlled by one
% parameter $\beta$.
The on-shell action, dual $CFT$ action and the source term can then be
derived using this massive self-dual condition as,
\begin{eqnarray}
\label{MASSIVE SELFDUAL ONSHELL ACTION}
I^{MS}_{os}[A^{a(0)}_{i,q}]&=&\frac{1}{2}(\beta-1)\int d^3p
\left[mA^{a(0)}_{i,p}A^{a(0)}_{i,-p}+\epsilon_{ijk}\left(A^{a(0)}_{i,p}
(ip_j)A^{a(0)}_{k,-p} \right.\right.\\ \nonumber
&-&\left.\left.\frac{1}{3}g\epsilon^{abc}\int d^3 l d^3 p 
A^{a(0)}_{i,q}A^{b(0)}_{j,l}A^{c(0)}_{k,p}\delta^{3}(q+l+p) \right)\right], \\
\Gamma^{MS}[A^{a(0)}_{i,q}]&=&-\frac{1}{2}(\beta-1)\int d^3p\left[m
A^{a(0)}_{i,p}A^{a(0)}_{i,-p}+\epsilon_{ijk}\left(A^{a(0)}_{i,p}(ip_j)
A^{a(0)}_{k,-p} \right.\right.\\ \nonumber
&-&\left.\left.\frac{1}{6}g\epsilon^{abc}\int d^3 l d^3 p
    A^{a(0)}_{i,q}A^{b(0)}_{j,l}A^{c(0)}_{k,p}\delta^{3}(q+l+p) \right)\right]. \\
{\rm and\ }J_{MS}&=&(\beta-1)\left[mA^{a(0)}_{i,-p}
+\epsilon_{ijk}\left(ip_jA^{a(0)}_{k,-p}
\right.\right.\\ \nonumber &-&\frac{1}{4}\left.\left.
g\epsilon^{abc}\int d^3 l d^3 p A^{b(0)}_{j,l}A^{c(0)}_{k,p}
\delta^{3}(q+l+p) \right)\right],
\end{eqnarray}
respectively. The on-shell effective action (\ref{MASSIVE SELFDUAL
  ONSHELL ACTION}) turns out to be proportional to the non-abelian
Chern-Simon action.

\section{``Approximate'' Electric-Magnetic Duality in SU(2)
  Yang-Mills in $AdS_4$}
\label{em-duality}\setcounter{equation}{0}

%In this section, we discuss Electric-Magnetic Duality in SU(2) Yang-Mills in $AdS_4$.
It is well-known fact that explicit electric-magnetic duality cannot
be demonstrated for non-abelian gauge field theory, pure U(1) gauge
theory equations of motion, on the other hand, are manifestly invariant.
Exchanging electric and magnetic fields is possible even for
Yang-Mills, but such a transformation is not a canonical
transformation\footnote{There is, in fact, a no-go theorem for this
  duality.  At least in a particular gauge this has been demonstrated
  in \cite{Deser3}.}.
%% In fact, this exchanging the fields 
%% turn out to be ``NOT'' canonical transformation

There is, however, an attempt to construct a canonical transformation
in SU(2) Yang-Mills, which is gives rise to an {\it approximate}
electric-magnetic duality transformation, if one restrict to
Yang-Mills action truncated upto cubic order interactions in weak
field expansion\cite{Deser2}.

To see this more clearly, let us explain the meaning of ``{\it
  approximate} '' electric-magnetic duality.  The authors in
\cite{Deser2} construct an infinitesimal canonical transformation
which is a natural extension of U(1) electric-magnetic duality to
SU(2) Yang-Mills, which is manifest symmetry in Yang-Mills action when
the action only retains cubic order interactions in small amplitudes
of gauge fields in it(i.e. they do not keep quartic order
interactions). Therefore, if Yang-Mills coupling vanishes, then this
symmetry becomes the usual duality in U(1).  However, this is not
precisely electric-magnetic duality in SU(2) Yang-Mills since the
variation of electric field is not proportional to magnetic field even
upto such a truncation. Therefore by ``approximate'' duality we mean that
there exists a canonical transformation which is the most natural
generalization of electric-magnetic duality in U(1).  
%% Another point that we need to mention is that they choose a particular 
%% gauge for Yang-Mills field: transverse gauge.  
It is worth mentioning at this point that this has been demonstrated
in a particular gauge for the Yang-Mills fields, namely the transverse
gauge.  In this gauge, components of gauge fields
surviving in the action are all transverse.  In any other gauge, the
transformation may be difficult to implement.
%% or the action is not to be
%% manifest to be invariant under it.

In this section, we will discuss ``{\it approximate} ''
electric-magnetic duality transformation for our system.  The
difference between flat space and $AdS$ space here only comes from
their boundaries.  In general, electric-magnetic duality is not a
manifest symmetry of the Lagrangian but it is a symmetry of equations
of motion. The total derivative terms in the Lagrangian which inhibit
this manifestation disappear in the flat space, if we suppose that all the
fields die off sufficiently fast at infinite boundary. However, Weyl transformed
action(\ref{flat-space-action}) from $AdS_4$ has conformal boundary
at $x^4\equiv r=0$ and gauge fields do not die off fast enough at this
boundary.  Therefore, we need to keep total derivative terms with
respect to `$r$'.

Now let us apply the canonical transformation of \cite{Deser2} to our
case. Yang-Mills action (\ref{flat-space-action}) can be written in terms
of the Legendre transform of the Hamiltonian as
\begin{equation}
\label{primitive-em-h-action}
S[A^a_i,E^a_i,A^a_r]= \int d^4 x \left(-E^a_i \partial_r A^a_i -\mathcal H[\Pi^a_i,A^a_i]\right),
\end{equation}
where the canonical momentum $\Pi^a_i=-E^a_i$, the  electric field, and the
Hamiltonian density $\mathcal H$ is given by 
\begin{equation}
\mathcal H=\frac{1}{2}\left(E^a_i E^a_i -B^a_i B^a_i\right) + A^a_r\left( \partial_i E^a_i + \epsilon^{abc}E^b_iA^c_i \right),
\end{equation}
where we set Yang-Mills coupling $g=1$ for convenience, $B^a_i=\frac{1}{2}\epsilon_{ijk}F^a_{jk}$, magnetic field, and negative sign in front of the magnetic field square in the
Hamiltonian density appears since we are working in the Euclidean
space .  Notice the
Legendre transform is taken with respect to the radial coordinate.
The gauge field component $A^a_r$ has no dynamics and in fact, it is a
Lagrange multiplier, which gives rise to the Gauss law constraint. By
imposing the Gauss law constraint, $(D_i E_i)^a=\partial_i E^a_i +
\epsilon^{abc}E^b_iA^c_i =0$, we can remove the terms which are proportional to $A^a_r$ from the action.
Another important point is the gauge choice. In \cite{Deser2}, authors
point out that it is very crucial to choose transverse gauge. Under
such a choice, longitudinal parts of electric fields becomes quartic
order in the weak field expansion in the action, so at the cubic
approximation we will not need to worry about those terms.  With all
these conditions, the action can be expressed as
\begin{equation}
S[E^a_i,A^a_i]=\int d^4 x \left[-E^{a,T}_i \partial_r A^a_i -\frac{1}{2}\left(E^{a,T}_i E^{a,T}_i -\bar B^a_i \bar B^a_i\right) 
- \frac{1}{2}\bar B^a_i \epsilon^{abc}\epsilon_{ijk} A^b_j A^c_k  + O((A^a_i)^\alpha (E^a_i)^\beta)\right],
\end{equation}
where $\alpha$ and $\beta$ are positive integers which satisfy an
inequality $\alpha + \beta \geq 4$ and $\bar
B^a_i\equiv\epsilon_{ijk}\partial_j A^a_k$. 
It turns out that this action is invariant upto cubic order
in small amplitude expansion of the fields under the following
infinitesimal transformation: 
\begin{eqnarray}
E^{a,T}_i &\rightarrow& E^{a,T}_i +\eta  \left(\bar B^a_i -\frac{3}{2}
  \epsilon^{abc}\epsilon_{ijk} (A^b_j A^c_k)^T\right) + {\rm \ higher\
  order,}\\ 
\label{longitudinal-em-transform}
E^{a,L}_i &\rightarrow& E^{a,L}_i
-\frac{1}{2}\eta\epsilon^{abc}\epsilon_{ijk}\left( A^b_jA^c_k -
  E^{b,T}_j\frac{1}{\nabla^2}E^{c,T}_k\right)^L+ {\rm \ higher\
  order,}\\ 
A^{a}_i &\rightarrow& A^{a}_i - \eta \frac{1}{\nabla^2}\epsilon_{ijk}\partial_j E^{a,T}_k + {\rm \ higher\ order,} \\ 
{\rm and\ } A^a_r &\rightarrow& A^a_r, 
\end{eqnarray}
where `higher order' denotes cubic or higher than cubic order in weak fields expansion, $\eta$ is the 
infinitesimal duality rotation angle and the superscripts $T$ and $L$
mean that only transverse and longitudinal parts of the terms would be kept
respectively.

Under such transformation, the action changes as
\begin{equation}
S[E^a_i,A^a_i] \rightarrow S[E^a_i,A^a_i] 
+ \frac{1}{2}\eta \int_{r=0} d^3 x\left( A^a \cdot \nabla \times A^a+\epsilon^{abc}A^a \cdot A^b \times A^c +E^a \cdot \frac{1}{\nabla^2}(\nabla \times E^a)\right),
\end{equation}
The action is invariant upto the boundary terms. These boundary terms
will be treated as an infinitesimal boundary deformations.  While the
last term in the boundary action is non-local, first two terms are
similar to the Chern-Simons term.  Since the duality transformation is
approximate, we are not able to get the relative factors correctly.
%% (We need to interpreted this, at least briefly!!)

One may wonder if transformation (\ref{longitudinal-em-transform}) of
the longitudinal part of the electric field will not be necessary since they appear at the quartic order and our transformations
are applicable only up to cubic part of the action.  However, the term
proportional to $A^a_r$ is eliminated from the action
(\ref{primitive-em-h-action}) by imposing the Gauss law constraint.
To ensure that the Gauss law constraint is not affected up to this
order requires the transformation  (\ref{longitudinal-em-transform}).

\section{Yang-Mills Instanton}
\label{Yang-Mills Instanton}\setcounter{equation}{0}
In the previous section, we have developed various kinds of
deformations to obtain the corresponding boundary actions. 
While there are many reasonable deformations, most of them are obtained
by doing small amplitude expansion about perturbative classical
solution.  
%% One of the clues for this question can be obtained by looking at
%% non-perturbative solutions in the bulk. 
In this section, we will consider a nonperturbative solution in the
bulk, namely, the instanton solution, and construct the boundary
action corresponding to this Yang-Mills instanton solution in the
bulk. The instanton solution in the flat space is known for a long
time and we use the same solution\cite{Belavin1,Rajaraman1,Stefan1}.
The reason is that in four dimensions Yang-Mills theory is classically
conformally invariant and the four dimensional anti-de Sitter space,
$AdS_4$, is conformally flat.  As a result the instanton solution to
Yang-Mills theory in the Euclidean $AdS_4$ has same form as that in
the $\mathbb R^4$.  There is a crucial difference between these two
cases because the Euclidean $AdS_4$ is conformally equivalent to
$\mathbb R^4_+$ because of the semi-infinite range of the radial
coordinate.  This fact plays an important role in determining the
boundary action.  We start our discussion with the 't Hooft
instanton\cite{'tHooft:1976fv} with winding number 1 which is a
solution to the self-duality condition (\ref{self-dual-condition}),
and is given by
 \begin{equation}
\label{the instanton solution}
 A^{a}_{M}(x,x_0,\rho)=-\frac{2}{g}\frac{\eta^{a}_{MN}(x-x_{0})^N}
{(x-x_0)^2+\rho^2},
 \end{equation}
 where, $\rho$ is a real parameter which is size of the instanton and
 $x^M_0$ indicates its position. For simplicity, we choose $x^4_0=0$, then
our instanton solution is located on $AdS$ boundary. The gauge condition is chosen as
 $\partial_{M} A^{aM}=0$ and $\eta^a_{ij}=\epsilon^{aij}$,
 $\eta^a_{ir}=-\eta^a_{ri}=\delta^a_{i}$ for $i=1,2,3$.  Even if
 equations of motion are the same under the Weyl rescaling defined at
 the beginning in the Sec.\ref{Bulk Solutions}, the gauge condition
 does not. Lorentz gauge condition $\partial_{M} A^{aM}=0$ in flat
 space is different from that in $AdS$ space, which is $\partial_{N}
 (\sqrt{-G}G^{NM}(r)A^{a}_M)=0$. However, the radial gauge is the same
 in both cases.  It is therefore convenient to work in the radial
 gauge (For details of gauge transformation and the radial gauge
 solution of Yang-Mills instanton, See Appendix.\ref{Yang-Mills
   Instantons in Radial Gauge}).

The field strength of 't Hooft instanton solution is given by
 \begin{equation}
\label{the instanton field strength}
 F^a_{MN}=\frac{4}{g}\eta^{a}_{MN}\frac{\rho^2}{((x-x_0)^2+\rho^2)^2},
 \end{equation}
 and the action has a finite value as
 $S[A_{instanton}]=\frac{8\pi^2}{g^2}$.

 In flat $\mathbb{R}^4_+$, the instanton solution(\ref{the instanton
   solution}) approaches pure gauge solution and the field
 strength(\ref{the instanton field strength}) becomes zero at $x_4=r
 \rightarrow \infty$. This region, however, gets mapped to the
 Poincare horizon in $AdS_4$ space under the Weyl scaling (See the
 beginning in Sec.\ref{Bulk Solutions}).  Therefore, Yang-Mills
 instanton solutions do not change the boundary conditions at the
 horizon. Interestingly, the instanton solution does not become pure gauge
 solution at $r=0$ and the field strength has the finite value.  As
 shown in Appendix \ref{Yang-Mills Instantons in Radial Gauge}, the
 Fefferman-Graham expansion of Yang-Mills instanton in the radial gauge
 near $AdS$ boundary is given by
\begin{equation}
\label{Fefferman-Graham expansion of instantons}
A^a_{i}=A^{a(0)}_{i}+r A^{a(1)}_{i}+O(r^2),
\end{equation}
with,
\begin{eqnarray}
\label{boundary Yang-Mills A1}
A^{a(0)}_{i}&=&-\frac{2}{g}\frac{\eta^a_{ij}(y-y_0)_j}{{(y-y_0)^2+\rho^2}}, \\
\label{boundary Yang-Mills A2}
A^{a(1)}_{i}&=&-\frac{4}{g}\frac{\delta^a_i\rho^2}{((y-y_0)^2+\rho^2)^2},
\end{eqnarray}
where we have defined a boundary coordinate $y^i \equiv x^i$, and
$y^2=\sum_{i=1}^3y^iy^i$.
% (The second one need to be double check and the others are to be
% refined.)  $A^{a(0)}_{i}$ should be an instanton solution of
% boundary effective action of the dual field theory with self-dual
% boundary condition. The boundary field $A^{a(0)}_{i}$ and its
% canonical momentum $A^{a(1)}_{i}$ satisfy small $r$ limit of
% self-duality condition on $AdS$ boundary, which is also given in
% Eq.(\ref{self-dual-in-boundary}). Under radial gauge, this becomes
$A^{a(1)}_{i}$ is related to $A^{a(0)}_{i}$ by the small $r$ limit of
the self-duality condition eq.(\ref{self-dual-in-boundary}), which in radial
gauge becomes
\begin{equation}
\label{self-dual-boundary-in-radial-gauge}
A^{a(1)}_i=\frac{1}{2}\epsilon_{ijk}F^{a(0)}_{jk}.
\end{equation}
It is easy to see that the boundary values of Yang-Mills instanton
solution (\ref{boundary Yang-Mills A1}) and (\ref{boundary Yang-Mills
  A2}) satisfy the boundary condition
(\ref{self-dual-boundary-in-radial-gauge}).  We want to
write down this boundary condition in terms of the boundary field
$A^{a(0)}_i$ only.  
%% Because the only clue to figure out the correct
%% boundary condition is by looking at given instanton solution, there
%% are still many possibilities for the boundary conditions.  One of many
Although there are various ways of expressing this boundary condition,
we find representation of the Yang-Mills
gauge field corresponding to the 't Hooft instanton solution in terms
of a scalar function $\lambda(y)$ is most convenient for writing the
boundary term.

The usual ansatz for Yang-Mills instanton solution is given by
\begin{equation}
A^{a(0)}_i=\frac{1}{g}\eta^a_{ij}\partial_jln(\lambda(y)), {\rm \ where \ }
\lambda(y)=\frac{\rho^2}{(y-y_0)^2+\rho^2}.
\end{equation}
This equation can be inverted to write $\lambda(y)$ as a non-local
function of $A^{a(0)}_i$, 
\begin{equation}
\lambda(y)=e^{\frac{g}{2}\epsilon^{ai}_{\ \ j}\int^y_{y_0}A^{a(0)}_i(z) dz^j}.
\end{equation}
The $A^{a(1)}_i$ can be written in terms of $\lambda(y)$ as
\begin{equation}
A^{a(1)}_{i}= -\frac{4}{g}\delta^a_i \frac{\lambda^2(y)}{\rho^2}.
\end{equation}
With these, we can rewrite the self-dual boundary
condition (\ref{self-dual-boundary-in-radial-gauge}) in terms of
$A^{a(0)}_{i}$ only as
\begin{equation}
\label{Self-dual-boundary-with_A}
\frac{1}{2}\epsilon_{ijk}F^{a(0)}_{jk}(z)
= -\frac{4}{g\rho^2}\delta^a_i e^{g\int^z_{0}\epsilon^a_{ij}
A^{a(0)}_i(\tilde z)d\tilde z^j},
%\left( = -\frac{4}{g}\delta^a_i \frac{\lambda^2(y)}{\rho^2} \right).
\end{equation}
where $z \equiv y-y_0$.
% In the action level, $LHS$ of the equation comes from Chern Simons
% action. $RHS$ can come from a Wilson line type non-local deformation
% terms.

We can now ask if it is possible to write down the boundary on-shell
action such that the boundary condition
(\ref{Self-dual-boundary-with_A}) is the equation of motion of this
boundary action.
%% Now, let us suppose that this is the right boundary condition for the
%% winding number 1 instanton background in the bulk.  One could(at
%% least in principle) construct the boundary on-shell action to provide
%% eq.(\ref{Self-dual-boundary-with_A}) as a boundary condition at $r=0$.
It is easy to seen that the left hand side of
eq.(\ref{Self-dual-boundary-with_A}) comes from the non-abelian Chern
Simons action.  The right hand side contains line integration in the
exponent.  This line integral resembles the Wilson line, but it, in
fact, corresponds a non-local interaction.  Such a term in the
deformed boundary conditions cannot be obtained within the
perturbative approach. Moreover, in the case of multi-instanton
solution\cite{'tHooft:1976fv} in the bulk, the corresponding boundary
condition would continue to have this type of non-local, although its
precise form is different from eq.(\ref{Self-dual-boundary-with_A}).

To discuss this boundary condition in general, we promote
eq.(\ref{Self-dual-boundary-with_A}) to a general boundary condition
and treat $\rho$ as a parameter in the corresponding boundary
theory. The boundary on-shell action providing the boundary condition
then takes the following form:
\begin{equation}
I_{os}=a \int^{\infty}_{-\infty} d^3z\left[
\epsilon_{ijk}\left(A^{a(0)}_{i}(z)\partial_jA^{a(0)}_{k}(z) 
-\frac{1}{3}g\epsilon^{abc}A^{a(0)}_{i}(z)A^{b(0)}_{j}(z)A^{c(0)}_{k}
(z)\right)+ \frac{1}{\rho^2}\mathcal {L}_{NL}\right],
\end{equation}
where $a$ is a real constant and $\mathcal {L}_{NL}$ is a Lagrangian
providing the non-local term in eq.(\ref{Self-dual-boundary-with_A})
and we have pulled out $\rho$ dependence explicitly.  We now note that the
coupling $\frac{1}{\rho^2}$ explicitly breaks the scaling symmetry as
$z \rightarrow Lz$ and $A^{a(0)}_{i} \rightarrow \frac{1}{L}
A^{a(0)}_{i}$ and its scaling property shows that it is a relevant coupling.

Since, we promote $\rho$ as a parameter in the boundary on-shell
action, we can take a limit as $\rho \rightarrow \infty$. In this
limit, $\mathcal{L}_{NL}$ will relatively suppressed, and
% One interesting feature is that Lagrangian producing $RHS$ of the
% equation is proportional to $\frac{1}{\rho^2}$ explicitly. In the
% action level, one can take a limit as $\rho \rightarrow \infty$,
% then
the boundary theory becomes approximately pure Chern Simons theory.

\section{Conclusion}
\label{sec:conclusion}

We studied various boundary conditions for $SU(2)$ Yang-Mills theory
in $AdS_4$ background.  One of the motivation was to introduce
interactions in the boundary CFT.  Momentum dependent cubic
interactions in Yang-Mills theory lead to non-trivial interaction
terms in the boundary action both in cases of Dirichlet and Neumann
boundary conditions. We computed bulk Yang-Mills solutions to the first
subleading order to incorporate the leading effects of Yang-Mills interaction.
We found that in case of the Dirichlet boundary condition the boundary
propagator is proportional to $|\vec q|$, where $\vec q$ is three
dimensional momentum.  The cubic interaction has the form
\begin{equation}
  \Delta_{ijk}^{D,abc}(q,l,p) \sim ig\epsilon^{abc}\delta^{3}(q+p+l)
  \frac{(l-q)_k\delta_{ij}+(p-l)_i\delta_{jk}+(q-p)_j\delta_{ik}}{2(|q|+|p|+|l|)}.
\end{equation}
The Neumann boundary condition on the other hand has the propagator 
proportional to $1/|q|$ and the cubic interaction is 
\begin{equation}
\Delta_{ijk}^{N,abc}(q,l,p) \sim \frac{ \Delta_{ijk}^{D,abc}(q,l,p)}{|q||p||l|} .
\end{equation}
Another motivation was to study more interesting boundary conditions
like massive and self dual boundary conditions in the context of
non-abelian gauge theory.  While the massive boundary condition gives
rise to massive gauge theory on the boundary, the self dual boundary
condition takes the form of Bogomolnyi equation in the small $r$
expansion around the boundary.  The combined massive and self dual
condition gives massive Chern Simons gauge theory action on the
boundary.  Equations of motion derived from this action were studied
as self-duality conditions in odd dimensions\cite{Townsend1}.

We studied the effect of approximate electric-magnetic duality on
$SU(2)$ Yang-Mills theory defined on $AdS_4$ and resulting boundary
contribution.  Although the symmetry is not exact it seems to point
towards a Chern-Simon like term on the boundary in addition to a
non-local piece.  It would be interesting to explore effects of
duality on boundary conditions in the $AdS$ space, particularly in the
context of supersymmetric gauge theories.

We also studied instanton solution in $AdS_4$ with unit charge.  While
it was a straightforward generalization of the solution in $\mathbb
R^4$ due to conformal invariance of classical action and self-duality
condition, implication of the solution are quite interesting.  In
contrast to what happens in $AdS_5/CFT_4$ correspondence, where
D-instantons in $AdS_5$ do not modify the boundary condition, in
$AdS_4$ case, the Yang-Mills instanton becomes pure gauge on the
Poincare horizon and modifies the boundary condition on $AdS$
boundary.  We showed that the boundary action is the Chern Simons
action with a non-local deformation.  It would be interesting to
understand this non-local deformation better.

In this paper we concentrated only on
the gauge field sector, it would be interesting to combine it with
analysis of fermion boundary conditions \cite{Henneaux:1998ch}.  In
particular, it would be interesting to classify supersymmetric
boundary conditions\footnote{For some work along these lines see
\cite{Sebastian1}.}.  This analysis however is beyond the scope of this
work but we will address some of these issues in future.

\subsection*{Acknowledgments} {We would like to thank Stanley Deser,
  Ashoke Sen, Rajesh Gopakumar, Suvrat Raju, Sudhakar Panda for useful
  discussion. Especially, Jae-Hyuk Oh thanks his ${\mathcal W.J.}$ We
  would also like to thank anonymous referee for useful suggestions
  and providing pointers in those directions.  This work is partially
  supported by 11-R$\&$D-HRI-5.02-0304 at Harish-Chandra Research
  Institute, India. }

 \section*{Appendix}
\appendix
\section{Gauge Transformation}
\label{Gauge Transformation}\setcounter{equation}{0}
The most general form of non-abelian gauge transformation is given by
\begin{equation}
A^{a\prime}_M (x^P)\frac{\sigma^a}{2}= V(x^P)\left( A^a_{M}(x^P)
\frac{\sigma^a}{2}-\frac{i}{g} \partial_M\right) V^{-1}(x^P),
\end{equation}
where
\begin{equation}
V(x)=exp\left( -ig\phi^a \frac{\sigma^a}{2}\right)
\end{equation}
and $\sigma^a$ are Pauli matrices.  It turns out that the right gauge
transformation up to quadratic order in gauge fields or gauge
parameter is given by
\begin{equation}
\label{GAUGE TRANSFORM}
A^{a\prime}_{M} \rightarrow
A^{a}_{M}+\partial_{M}\phi^a-g\epsilon^{abc}
A^{b}_{M}\phi^c+\frac{1}{2}g\epsilon^{abc}\phi^b \partial_{M}\phi^c
+{\rm \ higher\ order},
\end{equation}
where $\phi^a$ is a gauge parameter which would be expanded as $\phi^a
= \varepsilon \bar \phi^a + \varepsilon^2 \tilde \phi^a
+O(\varepsilon^3)$ and ``$higher$'' means that the higher orders in
weak fields, $\phi^a$ or $A^a_M$ .  We evaluate this relation order by
order in $\varepsilon$ as
\begin{eqnarray}
\label{first-order-gauge-transform}
{\rm\ First\ order\ in\ \varepsilon\ : \ }\bar A^{a\prime}_{M}
&\rightarrow& \bar A^{a}_{M}+\partial_{M} \bar \phi^a \\
\label{second-order-gauge-transform}
{\rm\ Second\ order\ in\ \varepsilon\ : \ }\tilde A^{a\prime}_{M} 
&\rightarrow& \tilde A^{a}_{M}+\partial_{M} \tilde \phi^a-g
\epsilon^{abc}\bar A^b_M \bar \phi^c
+\frac{1}{2}g\epsilon^{abc}\bar \phi^b \partial_M \bar \phi^c.
\end{eqnarray}
Under these transformation, the field strengths are transformed as
\begin{eqnarray}
{\rm\ First\ order\ in\ \varepsilon\ : \ }\bar F^{a \prime}_{MN} 
&\rightarrow& \bar F^a_{MN}, \\
{\rm\ Second\ order\ in\ \varepsilon\ : \ } \tilde F^{a \prime}_{MN} 
&\rightarrow& \tilde F^a_{MN}-g\varepsilon^{abc}\bar \phi^c \bar F^b_{MN}
%(\partial_M \bar A^b_N -\partial_N \bar A^b_M)
\end{eqnarray}

\section{Evaluation of the Second order Bulk Solution}
\label{Evaluation of the Second order Bulk Solution}\setcounter{equation}{0}
We start with eq.(\ref{N=r the second order Bulk equation}).  We plug
the first order solution(\ref{first order solution}) into this and we
get
\begin{eqnarray}
0&=& \left( \nabla^2\tilde A^a_{r}-\partial_r \partial_j \tilde A^a_j 
\right) -g\epsilon^{abc}\left( \nabla^2 \bar \phi^b \partial_r \bar 
\phi^c + (\partial_j \bar \phi^b + \bar A^{b
  T}_j)\partial_j \partial_r \bar \phi^{c}\right. \\ \nonumber
&+&\left.(\partial_j \bar \phi^c + \bar A^{c T}_{j})\partial_r 
\bar A^{b T}_{j}\right).
\end{eqnarray}
We want to solve this equation in momentum space, so we perform
Fourier transform for any fields appearing in the equation using
eq.(\ref{Fourier Transform}).  Then, the equation becomes
\begin{equation}
\label{r-equation}
q^2 \tilde A^a_{r,q}-iq_j \partial_r \tilde A^a
_{j,q}=-g\epsilon^{abc}
\int^{\infty}_{-\infty} d^3 p \left(  -p_j q_j \bar
  \phi^b_p \partial_r \bar\phi^c _{q-p}+\bar A^{cT}_{jp}\partial_r 
\bar A^{bT}_{j,{q-p}}-ip_j \partial_r (\bar \phi^c_p \bar 
A^{bT}_{j,q-p})\right).
\end{equation}
We would like to define $RHS$ of this equation as source terms, which
come from the first order solution.  Under radial gauge(\ref{the
  radial gauge}), the first terms in both side on the above equation
vanish since the gauge parameter $\bar \phi$ does not depend on $r$
and $\tilde A^a_r=0$.  Using
\begin{equation}
\partial_r \bar A^a_{M,q}(r)=-|q| \bar A^a_{M,q}(r),
\end{equation}
we get eq.(\ref{the final form of longitudinal}).

To manipulate $N=i$ equations, we substitute the first order
solution(\ref{first order solution}) into eq.(\ref{N=i the second order
  Bulk equation}). Then, we obtain
\begin{eqnarray}
0&=&(\partial_r^2+\nabla^2)\tilde A^a_i -\partial_i (\partial_r 
\tilde A^a_r+\partial_j \tilde A^a_j)-g\epsilon^{abc}\left\{ 
\partial_r (\partial_r \bar\phi^b (\partial_i \bar \phi^c + 
\bar A^{cT}_i)) \right. \\ \nonumber
&+&\left.\partial_j\left( (\partial_j \bar \phi^b + \bar A^{bT}_j)
(\partial_i \bar \phi^c+ \bar A^{cT}_i)\right)-(\partial_j \bar
A^{bT}_i-\partial_i \bar A^{bT}_j)(\partial_j \bar \phi^c + 
\bar A^{cT}_j)-\partial_r \bar \phi^c \partial_r \bar A^{bT}_i\right\}.
\end{eqnarray}
By performing Fourier transform, the momentum space expression of
the equation becomes
\begin{eqnarray}
\label{i-equation}
0&=&(\partial_r^2-q^2)\tilde A^a_{iq} +iq_i (\partial_r \tilde
A^a_{rq}-iq_j \tilde A^a_{jq})-g\epsilon^{abc}\int^{\infty}_{-\infty}
\left\{ \partial_r (\partial_r \bar \phi^b_p(-i(q-p)_i \bar
  \phi^c_{q-p}
+\bar A^{cT}_{i,q-p}))\right. \\ \nonumber
&-&iq_j(-ip_j \bar\phi^b_{p}+\bar
A^{bT}_{j,p})(-i(q-p)_i\bar\phi^c_{q-p}+\bar
A^{cT}_{i,q-p})-\partial_r \bar \phi^c_p \partial_r 
\bar A^{bT}_{i,q-p} \\ \nonumber
&-&\left. ( -ip_j \bar \phi^c_{p}+\bar A^{cT}_{j,p}  )(-i(q-p)_j \bar 
A^{bT}_{i,q-p}+i(q-p)_i \bar A^{bT}_{j,q-p})\right \}d^3 p.
\end{eqnarray}
To obtain solutions, we plug eq.(\ref{r-equation}) into
eq.(\ref{i-equation}) and we get
\begin{eqnarray}
\label{2nd order transverse solution}
(\partial^2_r-q^2)P_{ij}(q)\tilde A^a _{jq}&=& g \epsilon^{abc}
\int^{\infty}_{-\infty}d^3p\left\{ iP_{ij}(q)p_j\partial_r (\bar 
\phi^c_{q-p}\partial_r \bar \phi^b_p) +P_{ij}(q)\partial^2_r(\bar 
\phi^b_p \bar A^{cT}_{i,q-p}) \right. \\ \nonumber
&+&\frac{i}{2}q_j(p_jq_i-p_iq_j)\bar \phi^b_p \bar \phi^c_{q-p} +\bar 
\phi^c_p \left( (\partial^2_r+(2q-p)_k p_k )\delta_{ij}-q_iq_j \right)
\bar A^{bT}_{j,q-p} \\ \nonumber
&+& \left. \bar A^{cT}_{k,p}\left( \frac{iq_i}{q^2}\delta_{jk}
\partial^2_r-i(q-p)_i \delta_{jk}+iq_k\delta_{ij}-iq_j\delta_{ik}
\right)\bar A^{bT}_{j,q-p} \right\},
\end{eqnarray}
where $P_{ij}{(q)}=\delta_{ij}-\frac{q_iq_j}{q^2}$ is a projection
operator to the transverse part of gauge field.  Contracting $q_i$ to
both sides of the equation, one can see that both sides are identically
zero.  The radial gauge condition eliminates the first term on the
$RHS$.  The terms proportional to $\bar \phi^b \bar A^c_i$ are
combined and they are
\begin{equation}
\bar \phi^b \bar A^c_i \sim (|q-p|^2-q^2)P_{ij}(q)\bar \phi^b_p 
\bar A^c_{i,q-p},
\end{equation}
using $\partial^2_r A^a_{i,q}(r)=|q|^2 A^a_{i,q}(r)$. The term
proportional to $\bar \phi^b \bar \phi^c $ can be written as
\begin{equation}
\frac{i}{2}q_j(p_jq_i-p_iq_j)\bar \phi^b_p \bar \phi^c_{q-p} 
=-\frac{i}{2}q^2P_{ij}(q)p_j \bar \phi^b_p \bar \phi^c_{q-p}.
\end{equation}
These equations can be used to obtain eq.(\ref{the final form of TRAnsverse}).

\section{Bulk Solutions in the Position Space}
\setcounter{equation}{0} In Sec.\ref{Bulk Solutions}, we have obtained
bulk solutions in the momentum space.  In this section, we would provide 
position space expressions, which are given by
\begin{eqnarray}
\bar A^a_{i}(r,x)&=&\partial_i \bar\phi^a(r,x) +\bar A^{aT}_i(r,x) 
{, \  \  } \bar A^{a}_r=\partial_r \bar\phi^a(r,x){, \  \  } 
\partial_i \bar A^{aT}_i(r,x)=0, \\ \nonumber
{\rm \ and \ } \bar A^{aT}_{i,p}(r,x)&=&\cosh(\sqrt{-\nabla^2}r)\bar 
A^{aT(0)}_{i}(x)+\frac{1}{\sqrt{-\nabla^2}}\sinh(\sqrt{-\nabla^2}r)
\bar A^{aT(1)}_{i}(x),
\end{eqnarray}
with 
\begin{equation}
\bar A^{aT}_{i}(r,x)=e^{-\sqrt{-\nabla^2}r}\bar A^{aT(0)}_{i}(x),
\end{equation}
by the regularity condition.
To maintain the radial gauge, we need
\begin{equation}
\bar\phi^a(r,x^i) \rightarrow \bar\phi^a(x^i).
\end{equation}
The position space expressions for solutions up to $O(\varepsilon^2)$,
eq.(\ref{the transverse the secon order}) and eq.(\ref{the longitudinal
  the secon order}), are given by
\begin{eqnarray}
\tilde A^{aT}_{i}(x)&=&g\epsilon^{abc}\bar A^{cT(0)}_k(x) \alpha_{ijk}
(\partial_l,\overleftarrow{\partial}_l)e^{-(\sqrt{-\overleftarrow{\nabla^2}}
+\sqrt{-\nabla^2}) r} \bar A^{cT(0)}_j(x),
\\
\tilde A^{aL}_{i}(x)&=&-g\epsilon^{abc}\frac{\partial_i}{\nabla^2}
\left(  \bar A^{cT(0)}_j(x)\frac{e^{-(\sqrt{-\overleftarrow{\nabla^2}}
+\sqrt{-\nabla^2}) r}}{\sqrt{-\overleftarrow{\nabla^2}}+\sqrt{-\nabla^2}}
\sqrt{-\nabla^2} \bar A^{bT(0)}_j(x)  \right),
\end{eqnarray}
where
\begin{equation}
\alpha_{ijk}(\partial_l,\overleftarrow{\partial}_l)=\frac{\left(\left( 
\frac{\overleftarrow{\partial_i}+\partial_i}{(\overleftarrow{\partial_i}
+\partial_i)^2} \nabla^2+ \partial_i\right)\delta_{jk}
-(\overleftarrow{\partial}_k+\partial_k)\delta_{ij}+
(\overleftarrow{\partial}_j+\partial_j)\delta_{ik}\right)}
{(\sqrt{-\overleftarrow{\nabla^2}}
+\sqrt{-\nabla^2})^2+(\overleftarrow{\partial_i}+\partial_i)^2}
,
\end{equation}
and the differential operators with arrows indicate that such
operators act to the left and the operators without arrows would act
to the right.

\section{Yang-Mills Instantons in Radial Gauge}
\label{Yang-Mills Instantons in Radial Gauge}\setcounter{equation}{0}
The usual Yang-Mills instanton solution is given by
\begin{equation}
 A^{a}_{M}(x,x_0,\rho)=-\frac{2}{g}\frac{\eta^{a}_{MN}(x-x_{0})^N}
{(x-x_0)^2+\rho^2},
 \end{equation}
 For the further use, we need to transform this expression into radial
 gauge. In this section, we explicitly construct the gauge
 transformation from the above expression to radial gauge
 solution. First, we separate the instanton solution into
 $r$-directional and $i$ directional pieces as
\begin{eqnarray}
A^a_r(r,y,\rho)&=&\frac{2}{g}\frac{\delta^a_i(y-y_0)_i}
{r^2+(y-y_0)^2+\rho^2}, \\
A^a_i(r,y,\rho)&=&-\frac{2}{g}\left(\frac{\delta^a_i r}{r^2+(y-y_0)^2+
\rho^2}+\frac{\eta^a_{ij}(y-y_0)_j}{r^2+(y-y_0)^2+\rho^2}\right),
\end{eqnarray}
where $x^4 \equiv r$ and $x^i \equiv y^i$ for $i=1,2,3$. For some
gauge transformation, we want to eliminate $A^a_r$. It turns out that
such gauge transformation is given by
\begin{equation}
V(x^P)=e^{-Z(x^P)},
\end{equation}
where
\begin{equation}
Z(x)=-\frac{i\sigma^a
  \delta^a_{i}(y-y_0)_i}{\sqrt{(y-y_0)^2+\rho^2}}tan^{-1}
\left(  \frac{r}{\sqrt{(y-y_0)^2+\rho^2}} \right).
\end{equation}
The question is that what is the form of $A^a_{i}$ in the radial
gauge.  It has a form of
% \begin{eqnarray}
% A^a_i(r,y,\rho)\frac{\sigma^a}{2}&=&-\frac{1}{g}V(x^P)\sigma^a 
% \left(\frac{\eta^a_{ir}r}{r^2+(y-y_0)^2+\rho^2}+\frac{\eta^a_{ij}
% (y-y_0)_j}{r^2+(y-y_0)^2+\rho^2} \right.
% \\ \nonumber
% &-&\left.\partial_i\left(
% \frac{\eta^a_{rj}(y-y_0)_j}{\sqrt{(y-y_0)^2
% +\rho^2}}tan^{-1}\left(  \frac{r}{\sqrt{(y-y_0)^2+\rho^2}} \right)
%    \right)  \right) V^{\dagger}(x^P)
% \end{eqnarray}
\begin{equation}
A^a_i(r,y,\rho)\frac{\sigma^a}{2}=e^{-Z(x^P)}\frac{\sigma^a}{g} Q^a_i 
e^{Z(x^P)},
\end{equation}
where
\begin{eqnarray}
Q^a_i&=&-\left(\frac{1}{\sqrt{(y-y_0)^2+\rho^2}}tan^{-1}(\frac{r}
{\sqrt{(y-y_0)^2+\rho^2}})+\frac{r}{r^2+(y-y_0)^2+\rho^2} \right)
\delta^a_j \Sigma_{ij} \\ \nonumber
&-&\frac{\eta^a_{ij}(y-y_0)_j}{r^2+(y-y_0)^2+\rho^2},
\end{eqnarray}
and
\begin{equation}
\Sigma_{ij}=\delta_{ij}-\frac{(y-y_0)_i (y-y_0)_j}{(y-y_0)^2+\rho^2}.
\end{equation}

Even if we do not obtain a compact form of the solution, we might get
the near boundary expansion of this instanton solution using the
boundary expansion of $V(x^P)$ as
\begin{equation}
V(x^P)=1+\frac{i\sigma^a\delta^a_{j}(y-y_0)_jr}{{(y-y_0)^2+\rho^2}}
+O(r^3).
\end{equation}

With this, one can expand $A^a_{i}$ near boundary as
\begin{equation}
A^a_{i}=A^{a(0)}_{i}+r A^{a(1)}_{i}+O(r^2),
\end{equation}
where
\begin{eqnarray}
A^{a(0)}_{i}&=&-\frac{2}{g}\frac{\eta^a_{ij}(y-y_0)_j}{{(y-y_0)^2+\rho^2}}, \\
A^{a(1)}_{i}&=&-\frac{4}{g}\frac{\delta^a_i\rho^2}{((y-y_0)^2+\rho^2)^2}.
\end{eqnarray}
%(The second one need to be double check and the others are to be refined.)
$A^{a(0)}_{i}$ should be an instanton solution of boundary effective
action of the dual field theory with self-dual boundary condition.

%% 
%% 
%% 
%% {\bf more references to be added}

\end{document}